\documentclass[12pt]{elsarticle}

\usepackage{amsmath, amssymb, amsthm, amssymb}
\usepackage[scaled=1]{couriers}
\usepackage[T1]{fontenc}
\usepackage[english,spanish,russian]{babel}
\usepackage{color}
\usepackage{calc}
\usepackage{longtable}
\usepackage{lscape}
\usepackage{supertabular}
\usepackage{gensymb}
\usepackage{MnSymbol} 
\usepackage{textcomp} 
\usepackage{wasysym}
\usepackage{cancel}
\usepackage{tikz}
\usepackage[english]{babel}
\usepackage{fontawesome5}
\usepackage[bottom]{footmisc}
\usepackage[english]{isodate}
\usepackage{incgraph}
\usepackage{lineno}
\usepackage{ucs}
\usepackage{textcomp}
\usepackage{tikz}
\usepackage{ulem}
\usepackage[utf8x]{inputenc}
\usepackage{wasysym}

\usepackage{natbib}
\hypersetup{ colorlinks=true, citecolor=red, linkcolor=blue, urlcolor=blue}
\hypersetup{colorlinks=true, linkcolor=blue, filecolor=blue, urlcolor=blue}
\definecolor{micolor}{rgb}{0.8, 0.12,0}

\makeatletter

\definecolor{green}{RGB}{35.445, 165.2, 0}

\newcommand{\abs}[1]{\lvert #1 \rvert}

\newcommand{\Com}{\mathbb{C}}

\newcommand{\comillas}[1]{\textquotedblleft{}{#1}\textquotedblright{}}

\newcommand{\e}{\text{\textit{\textbf{e}}}}

\newcommand{\G}[1]{\text{\textsf{\textbf{N}}}\left\lbrace   #1 \right\rbrace}
\newcommand{\Gaus}[1]{\boldsymbol{N}\left\lbrace{#1}\right\rbrace}

\newcommand{\given}{\;\mid\;}

\newcommand{\J}{i \!\! i} 

\newcommand{\lindep}{\perp \!\!\! \perp }

\newcommand{\lqqd}{\boldsymbol{\square}}

\newcommand{\mf}[1]{\mathfrak{#1}}
\newcommand{\ngr}[1]{\boldsymbol{{#1}}}

\newcommand{\nlindep}{\not{\lindep}}
\newcommand{\N}{\mathbb{N}}

\newcommand{\Q}{\mathbb{Q}}
\newcommand{\RefDS}{200 }

\newcommand{\R}{\mathbb{R}}
\newcommand{\scomillas}[1]{\textquoteleft{#1}\textquoteright{}}

\newcommand{\s}{\textasciiacute{}{s }}

\newcommand{\TEXs}{\textbf{4.8.4}} 
\newcommand{\then}{\text{\wasytherefore{\quad }}}

\newcommand{\var}[1]{{\text{var}\left[ {#1} \right] }}

\newcommand{\Z}{\mathbb{Z}}

\journal{}

\begin{document} 
	\selectlanguage{english}
	
\begin{frontmatter}	
	\title{
	Fractal dimension: Problems and traps of its  estimation.
} 
	
	\author{Carlos Sevcik\corref{cor} MD, PhD, Em. Prof. \\ Center Biophysics and Biochemistry, (IVIC), Caracas, Venezuela.}
	
	\cortext[cor]{
		{Current Address:Av. Paralelo 124, Ent 2A,Barcelona, PO Code 08015, Spain.\phone: +34 69766 84n02. \XBox:carlos.sevcik.s@gmail.com. \textbf{ORCID} Number: 0000-0003-3783-6541.}
	}
	
	\begin{abstract}
		This chapter deals with error and uncertainty in data. Treats their  measuring methods and meaning. It shows that uncertainty is a natural property of many data sets. Uncertainty is fundamental for the survival os living species, Uncertainty of the \comillas{chaos} type occurs in many systems, is fundamental to understand these systems. 
	\end{abstract}
	
\end{frontmatter}


\section{Introduction.}

\subsection{Complexity as a fundamental problem of science.}

\textit{Complexity} is a fundamental characteristic of Our (The?) Universe. Without \textit{complexity} life will not exist \cite{HowManyUniverses2022}. 

Defining \textit{complexity} is one of the most challenging problems in science \cite{Hossenfelder2023a}, some scientists even consider it the most importan problem in science \cite{Hossenfelder2023c},\textit{ a problem which hinders all modern science}. Complexity is related to entropy, but is a condition between zero entropy and maximum (infinite?) entropy. The problem was first expessed by Maxwell as a \comillas{daemon}, \cite{Knot1911} and remained unsowed until \cite{Szilard1929} associated complexity and information. Szilwrd\s approach leas to the definition of information as \textit{negentropy} \cite{Shannon1948, Shannon1949} and \textit{algorithmic information} \cite{Chaitin1969, Chaitin1976, Chaitin1987}. Fractal ) dimension ($D$ is also a measure of complexity.

\subsection{Complexity, uncertainty and experimental error.}

The unpredictable data component of a random process studied under constant conditions, may also be called noise. The noise definition is particularly important when a random data sequence is studied, the so called time series. Alterations of conditions where the random may naturally occur, but it may also stem from error when recording data due to instrument failures or due to experimenters mistakes, this is the case of real error. 

A common view is to consider data dispersion as \scomillas{errors}, something related with experimenters or equipment \scomillas{mistakes}, which certainly do occur sometimes. Yet, an extremely important source of data uncertainty, is not \scomillas{error}, but a property of the system under study. Such is the case of quantum physics \cite{Heisenberg1927} and related disciplines. But fuzziness of data is fundamental importance in biology, if all elements of any population are equal, they can by exterminated by noxae such as a pandemic, or by defects resulting from inbreeding in a small uniform population. When a species is reduced to small sets of individuals, the species in condemned to extinction. Many fundamental biological signals (hearth beat, brain signals) in healthy individuals are fuzzy and chaotic, and brcome less chaotic in pathological states like heart beat \cite{Hopkins2005, Hopkins2006, Mayor2023} or 
electroencephalographic (EEG) signals \cite{Shamsi2021, Spacic2011, klonowski2003, Gob2005, Klonowski2005, Liu2005, Spasic2011a, Harne2014, Cukic2018, Perez2022, Hadiyoso2023} (a non exhaustive list of examples). Fractal dimension is also used to analyze lung sounds \cite{Gnitecki2004, Gnitecki2005, Hadjileontiadis2005}. Besides biomedical uncertainty, there is a large number of papers using fractal dimension in subjects such as geomagnetic field studies \cite{Gotoh2003}, mammary \cite{Pe2003}, ultrasound studies \cite{Chen2005} and machines and materials failures \cite{Grzesik2009, ValtierraRodriguez2019} and other situations \cite{DSuze2015a, Sharma2013, RodriguezHernandez2022}. Despite the abundance of publications where fractal dimension is calculated, the authors often do not try to understand the mathematics.

In this review we will consider several proposed modes to calculate the fractal dimension \cite{Higuchi1988, Higuchi1990, Katz1988, Sevcik1998a, Sevcik2010}, and to present several examples of \scomillas{dispersion} related to chaotic nonlinear systems which are some sources of fractality. We will also consider the Hurst\s coefficient \cite{Hurst1951} for which a simple relationship whih $D$ was proposed \cite{Mandelbrot1983, Hastings1993}, a relation currently considered wrong \cite{Mandelbrot2002, Gneiting2004, Gneiting2011, Sutcliffe2016}.

\subsection{Relevance of chaos and dynamical systems.}

Complex systems are often called nonlinear systems, strongly sensitive to initial conditions, systems that are also known as \scomillas{dynamical systems}. In these SYSTEMS, there is no randomness, but they change as a result of the accuracy of the calculations or due to very small environmental variations., The classical  example is earth climate shown by Loeenz \cite{Lorenz1963, Lorenz1969}  to be unpredictable, sometimes even at short lapses of time \cite{Lorenc1998, Dunne2013, Bozeman2019}.

Lorence \cite{Lorenz1963, Palmer2022}, proved that very small changes can produce very large whether changes, \scomillas{the flutter of a butterfly in China can produce a storm in America}, the so-called \textit{butterfly effect}, poetic name that is reinforced by the shape of the Lorenz attractor \cite{Lorenz1963} (See Figure \ref{F:Lorenz}). Dynamical systems are totally deterministic, not random, but they are unpredictable \cite{Lorenz1963, Burt1988, Lorenc1998, Jarraud1989}. 

Another form of uncertainty is called chaos. It is central to all fields of human knowledge including quantum physics \cite{Berry2001, Zurek20003, Hossenfelder2022j}. This chapter is an introduction to chaos and its difference from the statistical uncertainty to which the rest of this book relates. It is not a new concept started with Lorenz almost 60 years ago \cite{Lorenz1963} and remains central in almost all areas of scientific knowledge.

The impact of Loren butterfly effect on wheather prediction, has determined that the concept of chaos is usually said that was created by Lorenz \cite{Lorenz1963}, still the concept is earlier, it was introduced by Turing \cite{Turing1952} to explain the source of biological structure and complexity in biology, and by Belousov \cite{Belousov1959, Hudson1981, Zhabotinsky2007}.

Another result of chaos in nonlinear systems is turbulence \cite{Pavlos2012, Karakatsanisa2013}. A thin layer of ice on an airplaneLorenz\textquoteright{s} wings can cause enough turbulence to prevent it from flying. Chaos exists in vital functions such as the normal heart rate which is, within certain limits, chaotic, the absence of chaos in the heart rate indicates disease \cite{Pincus1991a, Goldberger1991, Iokibe1995, Pincus1991, Pincus2001, Hopkins2005, Pincus2006}, but if the chaoticity becomes extreme (the so-called ventricular fibrillation) it causes death. Examples of the significance of chaos are too many to cite here, this includes earthquakes \cite{Danos2006}, fluctuations in the stock market and the economy in general \cite{Mandelbrot1963, Mandelbrot1963, Mandelbrot1997, Kutner2003}, a couple of somewhat classic references are \cite{Mandelbrot1983, Hastings1993}.

\subsection{Lorenz\textquoteright{}s Uncertainty.}

But uncertainty is not just \comillas{error}. his was first shown in a log neglected model of wheather studied by Lorenz \cite{Lorenz1963} (S the system of Eqs. (\ref{E:Lorenz})). Lorenz\s system of equations was certainly sensitive to rounding errors, and the limited accuracy of the analog computer available to him. But the strange set of solutions (resembling a butterfly), although constrained to a definite volume of an Euclidean space,never crossed it self at a previous point (solution). The system was also unpredictable, it changed if was initialized with apparently similar values.

\begin{figure}[h!]
	\centering
	\includegraphics[width=6cm]{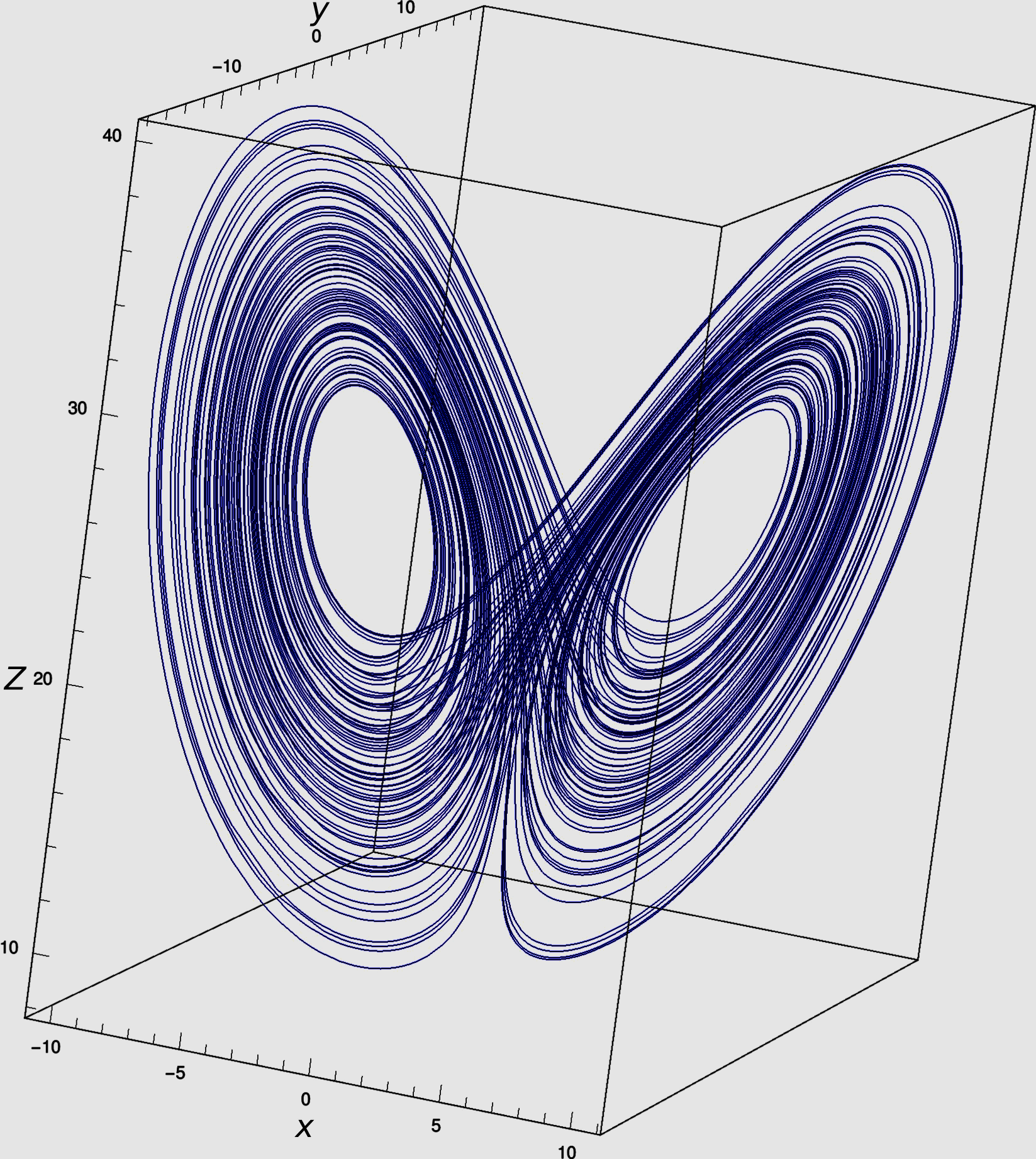}
	\caption{\footnotesize{\textbf{Lorenz\s attractor \cite{Lorenz1963}}. Latin letters \textit{x}, \textit{y} y \textit{z} represent an Euclidean three dimensional system coordinates. Example\s constant values are: $\sigma = 3$; $\rho = 26.5$; $\beta = 1$; initial coordinate\s values are $x_0 = -1$; $y_0 = 0$ y $z_0 = 1$. Figure presents $40000$ consecutive solutions of the system of equations solutions (\ref{E:Lorenz}). }}\label{F:Lorenz,pff}
\end{figure}

All possible values in a dynamic system usually usually exist in a finite space and are distributed in a region of that space called attractor. A classical example id Lorenz\s attractor \cite{Lorenz1963}, shown in Figure \ref{F:Lorenz} and represents the solutions of a simple set of differential equation built by Lorenz as a an atmospheric climate model. Lorenz\s equations are  \cite{Lorenz1963, Palmer2022, Lorenz1969}:
\begin{equation}\label{E:Lorenz}
	\begin{matrix}
		\cfrac{dx}{dt}&=&\sigma (y-x)  \qquad
		\cfrac{dy}{dt}&=&\rho x - xz-y \qquad
		\cfrac{dz}{dt}&=& xy - \beta z 
	\end{matrix}
\end{equation} 
where \textit{x}, \textit{y} and \textit{z} are the coordinates of an Euclidean system. Figure \ref{F:Lorenz} represents a series of solutions of this system. The trace in the figure represents a single continuous, which never crosses through the same, previous, point, and repents unpredictable displacements of which are solutions of the equation\s system. Perhaps the most important properties of the graphic, is that all solutions are confined in a subset (volume) of the Euclidean space called a strange attractor, Equation system (\ref{E:Lorenz}) was built by Lorenz as a climate model, is the first es el primer \textit{strange attractor} known and to describe climate (we also call it {whether}) and it showed that real whether (much more complex than the model) is unpredictable (except for short periods), changing the history of meteorology. Having a strange attractor is not unique to Lorenz\s system, many other systems have strange attractors too. 

Climate unpredictability has been well demonstrated after Lorenz. Wheather can only be predicted for short periods even with modern supercomputers of today, receiving information from sensors spread all over the world. Lorenz\s equations in laser models \cite{Haken1975}, electricity dynamos \cite{Knobloch1981}, convection loops \cite{Gorman1986},direct current motors without brushes \cite{Hemati1994}, electric circuits \cite{Cuomo1993} and chemical reactions \cite{Poland1993} among many more systems. Lorenz\textquoteright{}s-like systems are not this books\s object of study but they must be considered by its reader, since \textit{statistical methods are often applied to Lorenz-like dynamic systems, and are used to predict the properties of such systems, and the predictions failure is wrongly attributed to statistics and not to their unpredictable nature}. An example of this opinion studies, which when are made public, may (usually d) modify the public opinion that they \textit{objectively}  claim to study. In politics, it sometimes results it is common \scomillas{to bet on the winner} voting for or against him or her if a candidate produces fear. Classical examples of this effect is also the expectation produced by a medical treatment (which may lead to changing or selecting theproved that study subject or the experimenters performing the study), they could even modify the results observed (i.e. change the result). increasing de beneficial or adverse effects of the treatment studied, having or not having {faith} in the treatment could be another example. This is an extension of the quantum physics observer effect, extended to daily supra molecular. Lorenzian uncertainty was later called \textit{chaos}, and was the firs system known to contain \textit{chaotic} uncertainty.

\section{Time Series and the Fractal Dimension.}

\subsection{The concept of fractality.}

We define here as time series what, perhaps in better but longer common English, should be called: \textit{series of events which occur in time}. A set presented as a graph where the ordinate represents random events and the abscissa is the time at which each event occurs.

Studying systems, living or not, as dynamical (chaotic, as they are commonly called) nonlinear systems is of great interest in biology and medicine \cite{Elbert1994}. Fractal dimension analysis is a possible way to characterize dynamical systems and other complex curves and time series analysis is one of the most common ways used to calculate the fractal dimension from observables \cite{Elbert1994}. Time series analysis is also interesting \textit{per se}. However, this analysis can be related to complex concepts such as regularity, complexity or spatial extension \cite{Mandelbrot1983, Nicolis1989, Pincus1991}. A good example can be found in two series constructed by Pincus \textit{et al.} \cite{Pincus1991} to illustrate the complexities of heartbeat in healthy and diseased humans, these are:
\begin{equation*}
	90, 70, 90, 70, 90, 70, 90, 70, 90, 70, 90, 70, 90, 70, 90, 70,\ldots
\end{equation*}
and
\begin{equation*}
	90, 70, 70, 90, 90, 90, 70, 70, 90, 90, 70, 90, 70, 70, 90, 70,\ldots
\end{equation*}
The two series have the same mean and the same variance and the two values ($90\text{ or }70 $) have the same probability of occurring: \textonehalf. Rank statistics also do not distinguish between them. However, the two series are completely different; In the first we always know with total certainty which number follows once we observe an element of the series. In the second series, we only know that it will be $ 90 $ or $ 70 $, but our choice will be false in 50\% of cases.

The English term \textit{waveform}, which we translate here as \comillas{waveform} refers to the appearance of a wave much more complex than a \comillas{ripple} as defined by the Meriam Webstwe\s Dictionary, usually periodic \textit{versus} time. Any waveform is a series of points. Apart from classical statistical models such as statistical moments \cite{Guttman1971} and regression analysis, properties such as Kolmogorov-Sinai entropy \cite{Grassberger1983}, the apparent entropy  \cite{Pincus1991} and the fractal dimension \cite{Katz1988} have been proposed to address the problem of analyzing waveforms. The fractal dimension can provide information about spatial extension (tortuosity or its ability to fill space) and its self-similarity (ability to remain unchanged when the measurement scale changes), its self-affinity \cite{Barnsley1993}. Unfortunately, although there are rigorous methods for calculating the fractal dimension \cite{Grassberger1983a, Grassberger1983b, Badii1985, Ripoli1999}, their usefulness is severely limited since they demand great computing power and because their evaluation is time-consuming. In Euclidean space, waveforms are flat, two-dimensional curves.
\begin{quote}
	\textquotedblleft Clouds are not spheres, mountains are not cones, coastlines are not circles, and bark is not smooth, nor lightning travel in a straight line.  \cite{Mandelbrot1983}\textquotedblright
\end{quote}
Fractal geometry was introduced by Mandelbrot to describe natural forms. 
\begin{quote}
	\textquotedblleft fractal from the Latin adjective \textit{fractus}. The corresponding Latin verb \textit{frangere} "to break:" to create irregular fragments. $\ldots$ \textit{fractus} should also mean \comillas{irregular} \cite{Mandelbrot1983}\textquotedblright
\end{quote}
Nature is, above all is, \textit{complex}, 
\begin{quote}
	\textquotedblleft Nature exhibits not simply a higher degree but an altogether different level of complexity\cite{Mandelbrot1983}.\textquotedblright
\end{quote}
According to Mandelbrot \cite[pg. 15 and Chap. 39]{Mandelbrot1983}: 
\begin{quote}
	\textquotedblleft A fractal is by definition a set for which the Hausdorff{-}Besicovitch dimension strictly exceeds the topological dimension. Every set with a non integer \textit{D} is a fractal.\textquotedblright
\end{quote}

\subsection{The Hausdorff--Besicovitch dimention.}

The Hausdorff--Besicovitch dimension ($ D_{HB} $) \cite{Hausdorff1918, Besicovitch1929} of a {metric space} (for which a metric dimension can be defined) of a set can be defined as \cite{Mandelbrot1983}:
\begin{equation}\label{E:D_HB}
	D_{HB}=-\underset{\epsilon \rightarrow 0}{\lim }{\frac{\ln [N(\epsilon )]}{\ln (\epsilon )}} = -\underset{\epsilon \rightarrow 0}{\lim } \log_{\epsilon}[N(\epsilon)]
\end{equation}
where $N(\epsilon)$ is the number of{open balls} of radius ${\epsilon}$ needed to cover the set. In a {metric space}, given any point $ X $, an open ball with scepter in $ X $ and radius $\epsilon$, is a set of all points $ x $ for which the distance between the points $ X $ and $ x $ is less than $ \epsilon $, that is: $ dist\left[ \abs{x-X} \right] $.

The term {waveform} applies to the form of a wave usually drawn as a value at an instant, of a periodic nature, versus time. A classic example is momentum statistics and regression analysis, properties such as the Kolmodorov-Sinai entropy \cite{Grassberger1983}, the apparent entropy \cite{Pincus1991} and the fractal dimension \cite{Sevcik1998a, Sevcik2010} have been proposed to perform waveform regression analysis. the fractal dimension can provide information on statistical extent (tortuosity or the ability to fill space) and similarity (the ability to remain unchanged when the measurement scale changes) and self-affinity \cite{Barnsley1993}. In processes occurring in two Euclidean dimensions, waveforms are curves with coordinates, $ x $ and $ y $, which \textit{usually have different units}.

\subsection{Waveforms with non integer $D_{HB}$.}

For a number of waveforms
\begin{equation*} 
	(D_{HB} \in \R) \land (D_{HB} \notin \Com) 
\end{equation*}  
where $\R$ is the ser of real numbers, an $\Com$ is the set of all real numbers. which does not necessarily mean that $ D_{HB}>1 $. There are also functions such as, for example, the call \scomillas{dust} or {Cantor set} \cite{Cantor1884, Darst1993, Dovgoshey2006, CantorSet2022, ListOfFractals2022} for which $ D_{HB} = \frac{log(2)}{log(3)}= 0.630929753571\ldots $. These waveforms correspond to infinite {disjoint sets of points} which, however, constitute a waveform \cite{ListOfFractals2022}.

In an Euclidean system on $n_E \in \Z$ dimensions\footnote{$\Z$ = set of all integers numbers; $\leftrightarrow$ = indicates a set of numbers limiters} 
\begin{equation}
	D_{HB} \in \{0 \leftrightarrow n_E\} \in \R
\end{equation}
which implies $\{0 < D_{HB} \leq 1\} $. In words, $D_hB$, or in general $D$ are real positive numbers, which may be smaller than $1$. This hoccurs in Cantor dets \cite{Cantor1884, CantorSet2022} also called \comillas{Cantor dudtd} in one dimensional Euclidean spaces,. 

\section{Estimators of fractal dimension.}

\subsection{The dimension of a waveform suggested by Katz \cite{Katz1988}.}

Fractal waveform analysis was initially proposed by Katz \cite{Katz1988}, who proposed that the complexity of a waveform can be represented by what Mandelbrot \cite{Mandelbrot1983} called the {fractal dimension}, and represented by Katz as $\Phi$ (represented as $D_K$ in this book). Katz \cite{Katz1988} said that the fractal dimension taking {$N$} samples measured empirically at constant intervals of the abscissa of the waveform.

Katz's equation \cite{Katz1988} was based on an observation by Mandelbrot \cite{Mandelbrot1983} where he pointed out that river causes were fractal structures where the fractal dimension could be approximated with an equation similar to Katz's, relating the length of the riverbed, with the greater distance separating two points of the river basin,. 

The procedure suggested by Katz \cite{Katz1988} discretizes the waveform producing $N'=N-1$ rectilinear segments from which with the notation, with the equation of Katz's \cite{Katz1988} below:
\begin{equation}\label{E:D_K}
	\Phi =\frac{\log (N')}{\log (N')+\log (d/L)} = D_K
\end{equation}
where $ d $ is the \textit{planar extension} of the \cite[Chapter 12]{Mandelbrot1983} curve and $ L $ is the length of the discretized curve defined as:
\begin{equation}
	\begin{split}
		d &= \max\left[ \text{dist}(i,j)\right] \\ 
		L &= \overset{N'}{\underset{i=0} \sum} {\text{dist}(i,i+1)}
	\end{split}
\end{equation}
where $ \max $ means the maximum $\text{dist}(i,j)$, of the distance the points $ i $ and $ j $ of the curve. For a curve that does not cross itself usually, but not always, $d=\max \left[ dist(1,i)\right] $.

\subsection{Sevcik\s fractal dimension \cite{Sevcik1998a, Sevcik2010}: a simple method to calculate the fractal dimension of waveforms.}

An expression to calculate the fractal dimension of a waveform is obtained from the  Hausdorff--Besicovitch ($ D_{HB} $) dimension \cite{Hausdorff1918, Besicovitch1929}. Mandelbrot\s {fractal} (see for example \cite{Mandelbrot1986}) is real number, not an integer. The Hausdorff-Besicovitch dimension \cite{Mandelbrot1986} of a metric space  can be expressed (for a very understandable discussion of metric spaces see Barnsley \cite{Barnsley1993}) as:
\begin{equation}\label{E:DimHausBes}
	D_{HB}=-\underset{\epsilon \rightarrow 0}{\lim }{  \frac{\ln [N(\epsilon )]}{\ln (\epsilon )} } = \underset{\epsilon \rightarrow 0}{\lim } \; \log_{\epsilon} \left[ N\left( \epsilon\right)  \right]
\end{equation}
where the notation is equal to that of Eq. (\ref{E:D_HB}). In a metric space given any point $ P $, an open ball with center at $ P $, $\epsilon$, is the set of all points $ x $ for which $ dist(P,x) < \epsilon $, at any length $ L $ can be divided into $N(\epsilon)=L/(2 \cdot \epsilon)$ long segments ${2 \cdot \epsilon}$, and any of them be covered by $ N$ open balls of radius $\epsilon$. Therefore Eq. (\ref{E:DimHausBes}) can be rewritten as
\begin{equation}\label{E:DimHausBes_2}
	\begin{split}
		D_{HB}&=\underset{\epsilon \rightarrow 0}{\lim }\left[\frac{-\ln (L)+\ln (2\cdot \epsilon )}{\ln (\epsilon )}\right]\\
		\cdots &=\underset{\epsilon \rightarrow 0}{\lim }\left[1-\frac{\ln (L)-\ln (2)}{\ln (\epsilon )}\right]\\
		\ngr{\then D_{HB}} &=\ngr{\underset{\epsilon \rightarrow 0}{\lim }\left[1-\frac{\ln (L)}{\ln (\epsilon )}\right] \quad \lqqd}
	\end{split}.
\end{equation}
Waveforms in two diminutions are sets of points with coordinates which may have distinct units. Since the topology of a metric space does not change under linear transformations, it is convenient to linearly transform one waveform into another in a normalized space, where all axes are equal and have legth $1$. This can be done with two linear transformations embedded into an equivalent metric space. The first transformation normalizes the curve\s abscissa as:
\begin{equation}\label{E:Sevcik_x_tranf}
	x_{i}^{\text{*}}=\frac{x_{i}}{x_{max}}.
\end{equation}
Where $ x_{i} $ are the original values of the abscissa, and $ x_{max} $ is the maximum ${x_i}$. The second transform normalizes the ordinate as follows:
\begin{equation}\label{E:Sevcik_y_tranf}
	y_{i}^{\text{*}}=\frac{y_{i}-y_{min}}{y_{max}-y_{min}}
\end{equation}
where ${y_{i}}$ are the original values of the ordinate, and $y_{min}$ and $y_{max}$ are the minimum and maximum $y_{i}$, respectively. 

The two linear transformations map $N$ points of the transform into another that belongs to a unit square. This square can be displayed as a grid of ${N \cdot N}$ cells. $ N $ of them containing a point of the transformed wave. A linear transformation applied to a function in a linear metric space does not alter the Calculating $ L $ of the transformed wave and Rolando $\epsilon=1/(2 \cdot N')$ the Eq. (\ref{E:DimHausBes_2}) is made 

\begin{equation}\label{E:D_S}
	\begin{split}
		D_{HB} &= \Phi \\
		\cdots &= \underset{N \rightarrow \infty}{\lim }\left( 1+\frac{\ln (L)}{\ln (2 \cdot N')} \right)\\
		\ngr{\then D_{HB} = \Phi}&\ngr{= \underset{N \rightarrow \infty}{\lim } \left(D_S\right) \qquad \lqqd}
	\end{split}
\end{equation}
the approximation to $\Phi$ expressed in Eq. (\ref{E:DimHausBes_2}) expressed in Eq. (\ref{E:D_S}), improves as $N' \rightarrow \infty$. The Eq. (\ref{E:D_S}) as simply $ D $, in \cite{Sevcik1998a, Sevcik2010} as an anonymous derivative of the Hausdorff-Besicovitch dimension \cite{Besicovitch1929, Hausdorff1918}, this fractal dimension estimator has been called with increasing frequency: \comillas{Sevcik\s} dimension \cite{Sharma2013, Diao2017, Shi2018, Nepiklonov2020, Xue2020, Kolodziej2020}. So since our work of 2022 \cite{RodriguezHernandez2022} we call it \comillas{Sevcik\s fractal dimension}, $D_S$.

\subsubsection{Approximation to the variance of $ D_{S} $.}

Although $\Phi$ is a topological invariant of a set and a metric space, $ D_S $ is only an empirical estimate of $\Phi$ with some uncertainty based on a set of points sampled from a wave; $ D_S $ is therefore a random variable. The ratio between $\Phi$ and $ D_S $ is similar to that between a mean of a \textit{population} ($ \mu $) and the mean $ \bar{x} $ estimated when sampling the population; Although $ \mu $ is a population invariant, $ \bar{x} $ will change with sampling. Just as $ \bar{x} $ converges towards $ \mu $ as the sample approaches population size, $ D_S $ converges towards $ \Phi $ as $ N'\rightarrow \infty $. Now we will derive an expression for $ var(D_S) $ (the variance of the $ D_S $) from the estimate of $ D_S $ obtained by mastering $ N' $ points of a wave. It should be obvious from the derivation of $ D_S $ and the non-stationary character of the values of $ D_S $ determined with Eq. (\ref{E:D_S}), $\var({D_S})$ does not provide information about the asymptotic value of $D_S$ obtained as ${N'\rightarrow \infty}$. The variance of $ D_S $ can be estimated starting from the following expression:
\begin{equation}\label{E:varDS}
	\begin{split}
		\var{D_S} &= \var{1+ \frac{\ln (L)}{\ln (2\cdot N') )}} \\
		\ngr{\then \var{D_S}}&=\ngr{ \var{\frac{\ln (L)}{\ln (2\cdot N\text{{\textquotesingle}})}}  \quad \lqqd}
	\end{split}
\end{equation}
The approximate solution to Eq. (\ref{E:varDS}) can be obtained by recalling that the variance any random variable set may be  $\{x_i\}$ approximated with a Taylor series (see for example \cite{Colquhoun1971})) as:
\begin{equation}\label{E:varSet_x}
	\begin{matrix}
		\var{x_{1},x_{2},\ldots ,x_{i},\ldots ,x_{k}} \\
		\cdots \approx \overset{k}{\underset{i=1}{\sum }}{\left[\left(\frac{\partial \left[f\left(x_{1,}x_{2},\ldots ,x_{i},\ldots ,x_{k}\right)\right]}{\partial x_{i}}\right)^{2}\cdot \text{var}(x_{i})\right]}\hfill
	\end{matrix}
\end{equation}
which for Eq. (\ref{E:varSet_x}) produces:
\begin{equation}\label{E:varSet_x2}
	\var{D_S} = \frac{\var{L}}{L^2 \cdot \ln(2 \cdot N')^2} 
\end{equation}
as $ L $ is the sum of $ N' $ segments of length $\Delta y$, Eq. (\ref{E:varSet_x2}) is equivalent to
\begin{equation*}\label{E:varD:SPrel}
	\var{D_S} = \frac{ N'
		\cdot \var {{\Delta (y)} } }{L^2 \cdot \ln (2 \cdot N')^2} 
\end{equation*}
where ${\text{var}(\Delta y)}$ may be estimated from the data as:
\begin{equation}\label{E:VarDelta_y}
	\var{\Delta y} = \dfrac{1}{N'} \sum_{i=1}^{N} \left( \Delta x_i - \overline{\Delta x} \right)^2 
\end{equation}
where ${\overline{{\Delta y}}}$ is the mean length of the segments. In this way combining Eqs. (\ref{E:varD:SPrel}) and (\ref{E:VarDelta_y}) we get: 
\begin{equation}\label{E:VarDelta_y_LN}
	\ngr{\then \var{D_S} =\dfrac{ \sum_{i=1}^{N} \left( \Delta x_i - \overline{\Delta x} \right)^2 }{L^2 \ln\left(2 \cdot N' \right)^2 } \qquad \lqqd}.
\end{equation}

\subsubsection{Convergence of $D_S$ towards $D_{HB}$, an analytic solution for Koch\s triadic curve.}\label{SS:Koch}\label{S:Convergence}

\begin{figure}[h!]
	\centering
	\includegraphics[width=10cm]{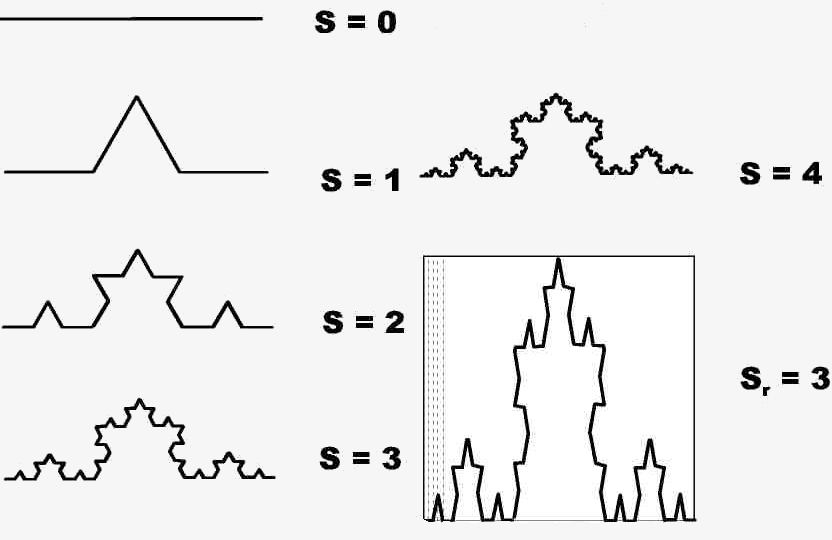}
	\caption{
		\footnotesize{
			\textbf{Characterization of the Koch curve.} The Koch curve is built on the three sides of an equilateral triangle, here I consider a single side to better visualize the convergence $D_S \rightarrow D_{HB}$ for the Koch\s snowflake. I start with a line of length $1$. Above and to the left of the graph (Stage 0 or $S = 0$, in the figure) a line of legth 1 ($\lambda = 1$) is presented. Then, at each stage, a segment of length $\frac{\lambda \cdot S}{3}$ is cut and moved up from each straight line, and an equal segment of length $\frac{\lambda \cdot S}{3}$ is added forming two sides of a new \comillas{triangle} lackin its \comillas{basal} segment. Repeating this an infinite number of times, one obtains Koch\s snowflake, each $S$ stage is indicated by the letter $ S = 0,1,2,\ldots,\infty$ Below and to the right of the figure, stage 3 ($ S=2 $) is shown scalded into a square of side $1$, in stage 3 ($ S_r=3 $), see the text for more details. This figure was adapted from Figure 4 of \cite{Sevcik1998a, Sevcik2010}.
		}
	}\label{F:CopoKoch}
\end{figure}

The asymptotic convergence of $ D_S $ to $ \Phi$ can be proven from the Koch triad. Although I derived the equation (\ref{E:D_S}) for waves, that is, plane curves that are sets of pairs of dots $(x_i, y_i)$ such as $x_i \rightarrow \infty$ when $i \rightarrow \infty$. However, on at least some occasions the utility of $ D_S $ $ \Phi$ extends to the field of waveforms, I will demonstrate that with the famous Koch triad shown in Figure \ref{F:CopoKoch}.

The following properties can be easily verified to be true for the triadic curve at any stage. \textit{S}:
\begin{equation}
	\begin{matrix}
		n_{s}=2^{2S} \qquad \qquad &L =\left[\frac{4}{3}\right]^{S} \\
		\\
		l_{s}=3^{-S} \qquad \qquad &n_{hs}=\frac{2^{2\cdot S}-1}{3}+1 \\
		\\
		n_{is}=2^{2\cdot S}-\left(\frac{2^{2\cdot S}-1}{3}+1\right)  \qquad \qquad &n_{v}= 2^{2\cdot S}-1 \\
		\\
		K_{h}=\frac{1}{\sqrt{12}} \qquad \qquad&
	\end{matrix}
\end{equation}
where: $ S $, is the number of the {stage} (as used in Figure \ref{F:CopoKoch}); $n_s$, is the number of segments that form a curve; $ L $, the length of the curve; $l_s$, is the length of each segment; ${n_{hs}}$, is the number of segments that have \comillas{horizontal} segments (that is, they can be extended as a line parallel to the line parallel to segment 0); ${n_{is}}$, is the number of \comillas{inclined} segments (This is non-horizontal as defined relative to ${n_{hs}}$; ${n_v}$ is the number vertex on the curve; ${K_h}$ is the height of the equilateral triangles constructed in stage 1, measured perpendicular to the line extending the horizontal segments at its base. This centers the open ball elements of radius.
\begin{equation*}
	\epsilon =\frac{3^{-S}}{2}
\end{equation*}
at both terminals of the triadic curve, and at each intersection of segments on the curve, therefore, we have
\begin{equation*}
	N(\epsilon )=1+2^{2 \cdot S}
\end{equation*}
of those open balls that are required to cover the curve. starting from the expression for ${\epsilon}$ and $N(\epsilon)$ and Eq. (\ref{E:D_HB}) we obtain the fractal dimension (Hausdorff{-}Besicovitch) of the curve as:
\begin{equation*}
	D_{HB}=\underset{S\rightarrow \infty }{\lim }-{\frac{\ln\left(1+2^{{2S}}\right)}{\ln \left(\frac{3 ^{-S} }{2} \right)}}=\frac{\ln(4)}{\ln (3)}=1.2618\ldots 
\end{equation*}
which is also the similarity and coverage dimension of the triadic curve. To test the capability of the Eq. (\ref{E:D_S}) to predict ${D_{HB}}$ in any case of the triadic curve we have to transform the curve as follows:
\begin{equation*}
	\begin{gathered}x_{i}^{\text{*}}=x_{i}\hfill\\y_{i}^{\text{*}}=y_{i}\cdot \sqrt{12}\hfill \end{gathered}
\end{equation*}
This is shown at the bottom of Figure \ref{F:CopoKoch} for the $3^S $ stage of the build process. The transformation does not modify the length of the horizontal components of the Koch curve but extends the length of all inclined sections that it becomes.
\begin{equation*}
	l_{is}= \sqrt{\frac{1}{6^{2 \cdot S}} + \frac{12}{6^{2 \cdot S}}} = \frac{\sqrt{13}}{6^{S}}.
\end{equation*}
And the curve at this stage becomes $S$ and we have
\begin{equation*}
	L=n_{hs} \cdot l_{s}+n_{s} \cdot l_{is}=\left(\frac{2^{2 \cdot S}-1}{3}+1\right)\cdot{\frac{1}{3^{S}}}+\left[2^{2 \cdot S}-\left(\frac{2^{2 \cdot S}-1}{3}+1\right)\right] \cdot {\frac{\sqrt{13}}{6^{S}}}
\end{equation*}
then, in order to provide that each vertex of the curve corresponds to a cell of the normalized square we have that
\begin{equation*}  
	N'=3^{S}
\end{equation*}
is outlined as dotted lines in Figure 4 ($s_r = $3). Replacing in Eq. (\ref{E:D_S})
\begin{equation}\label{E:Convergencia_D_Koch}
	\begin{split}
		D_{S} &= \underset{S\rightarrow \infty }{\lim } \left[1+\frac{\ln \left\{\left(\frac{2^{2\cdot S}-1}{3}+1\right) \cdot {\frac{1}{3^{{S}}}}+\left[2^{2 \cdot S}-\left(\frac{2^{2 \cdot S}-1}{3}+1\right)\right] \cdot {\frac{\sqrt{13}}{6^{S}}}\right\}}{\ln \left(2\cdot 3^{S}\right)}\right] \\
		\cdots &= \frac{\ln (4)}{\ln (3)} \\
		\ngr{\then D_{S}} &=\ngr{\frac{\ln (4)}{\ln (3)}  = D_{HB} = \Phi = 1.26185950714\ldots \qquad \lqqd} 
	\end{split}
\end{equation}
as it should be. \textbf{Thus, the limit of} $\ngr{D_S}$ \textbf{\textit{is indeed}} $\ngr{\Phi}$ when $\ngr{N \rightarrow \infty}$ \cite{Sevcik1998a, Sevcik2010}.

\subsubsection{Sevcik\s fractal dimension convergence to $ D_{HB} $.}

The main limitation of $ D_S $ is that the speed of convergence towards $ D_{HB} $, the latter is, generally speaking, unknown. Processes such as the one described by Eq. (\ref{E:Convergencia_D_Koch}) suggest that convergence occurs, but say little  about the speed of convergence. 

Part of the problem is that the original convergence presumes that in the set of pairs $ \{(x_i, \in \R) \; \ngr{\land}  \; (y_i \in \R) \}$, but this is not necessarily always true. Then: What happens, if we use expressions (\ref{E:D_S}) and (\ref{E:VarDelta_y_LN}) but $ \{(x_i, \notin \R) \; \ngr{\lor}  \; (y_i \notin \R) \}$?. This is the case when using $D_S$ to determine whether the series of digits of $ \pi $ is infinitely aperiodic , which in the field of aperiodic, non-rational numbers also called {normal} \textbf{(without, in this case, having any relation to the {Gauss pdf})} \cite{Sevcik2016, Sevcik2017b} : 
\begin{quote}
	The question has been asked by many authors \cite{Bailey2001, Bailey2002}. The definition of normal number \cite{Borel1909} is \cite[pg. 299]{Sierpinski1988}: Let $ g $ be a natural number $ > 1 $; we write a real number $ x = [x] + (0.c_1 c_2 c_3 . . .)_g $ as a decimal in the scale of $ g $. 
\end{quote}
For any digit $ c $ (on the scale $ g $) and each natural number $ n \in \N $, we denote $ I(c, n) $ the number of those digits rm the sequence $ c_1, c_2, \ldots, c_n $, which are equal to $ c $. Then
\begin{equation*}
	\lim_{n \to \infty} \left [\dfrac{I(c,n)}{n}\right]= \dfrac{1}{g}
\end{equation*}
For each of the $9$ values  $\neq 0$ of $c $, then the number $x $ called normal on the scale of $g $. A number that is natural on the $g $ scale is called {absolutely natural} \cite{Becher2002}. For a number with base $10$, the definition implies that $c_i \in \N [0, 9]$ if it must be true for any number $ x \in \R$ as $ N \rightarrow \infty $. Therefore, various authors use statistical tests such as comparing the frequencies of each digit in the sequences of the decimals of $ \pi $, and the frequencies of various combinations of decimal digits \cite{Bailey2012a}. This approach is unsatisfactory since all possible combinations cannot be evaluated, and is limited by the $ N $ values of other series studied, as well as by the size of sub samples of which $ N $ was selected to perform statistical tests.

The fractal analysis described \cite{Sevcik2016, Sevcik2017b} considered the set of decimals of $ \pi $, and calculated its approximate fractal dimension $D_S$ for $ N = 10^1, 10^2, 10^3, \ldots, 10^9 $ and showed that
\begin{equation*}
	\underset{N\rightarrow \infty }{\lim } D_{S} \approx 2
\end{equation*}
All series of type $ y_i = f(\{\theta_k\}_i) $, where $ f (\{\theta_k \}_i $) is some kind of random variable that depends on a set of parameters. This was found true in \cite{Sevcik2016, Sevcik2017b} for series of \textit{\textbf{independent}} real random numbers distributed with a pdf such as Gauss', Poisson's, exponential, or uniform $U_\R[0, 1]$, as well as discrete distributions such as $U_\Z [0, 9]$ for \textit{the decimal sequence of} $ \pi $. We have also seen that $ D_S $ for series by observing the condition represented as the equation\footnote{$\lindep$= linearly independent; $\forall$=for all;$\Rightarrow$=implies that}
\begin{equation}
	(y_i \lindep y_{k \neq i} ) \forall y_{k\neq i} \Rightarrow \underset{N\rightarrow \infty}{\lim} D_S = 2,
\end{equation}
is obtained under randomization. I.e.: $D_S$ does not change under randomization, \textit{\textbf{is a white noise}} \cite{Hastings1993}, and this is a property of the sequence of decimal digits of the sequence of digits of $ \pi $. Randomization increases the Boltzmann entropy (Section 4.2 and equation (4–30) of \cite{Boltzmann1964}) or the algorithmic type \cite{Sinai1959, Kolmogorov1964, Chaitin1969} of a random sequence; a sequence maximizes its entropy or equivalently, or has a maximum entropy or information content \cite{Szilard1929, Shannon1948}. The conclusions could be falsified by assuming that the singularity in the infinite series functions used to calculate the digits \cite{Chudnovsky2000} exist.

For more details on the decimals of $ \pi $ and its fractal dimension refer to the original work of Sevcik \cite{Sevcik2016, Sevcik2017b}, there. limited by the capacity of the available computers were calculated $ 10^9 $ of $ \pi $ decimal digits with the algorithm of Bellard \cite{Bellard2010}, already by that time the maximum number of decimal digits of $ \pi $ known was $ 2.7 \times 10^{12} $ digits calculated with the same algorithm \cite{Bellard2010}. At the time of writing this article $ 2,699,999,990,000 \approx 2,699 \ldots \times 10^{12}$ decimals of $ \pi $ are known, \cite{Pi2024}, store that number of decimals requires $ \approx 2.7 $ teraB of disk. But using $10^9 $ of decimals of $ \pi $ is enough to reach a value of $ D_S \approx 1.888421 \pm 10^{-6} $ while a sequence of of $ 10^9 $ decimal digits of type $ U_\Z(0.9) $ decimal has a $ D_S \approx 1.88743881 \pm 2\times10^{-6} $. 

Since we do not know the distribution of the decimals of $ \pi $, the approximation of the inequality of Vysochanskij–Petunin \cite{Vysochanskij1979, Vysochanskij1980, Vysochanskij1983} was used to compare the diverse values of $ D_S $ obtained when the decimals of $ \pi $ studied were increased by taking the first in the sequence of $ 10, 100, 1000, \ldots, 10^7, 10^8, 10^9 $ $\pi$ digits and a sequence of the same length of decimal distributed uniformly as $ U_\Z(0.9) $ or $ U_\R(0.1)) $ versus the randomized sequence of the same number of $ \pi $ digits. However, when $D_S$, was used to compare the sequence o $ \pi $ decimals with the same randomized sequence of $ \pi $ decimals and a sequence of the same length of $ U_\Z(0,9) $ decimals, no statistically significant differences were found ($ 0.11 < P < 0.46 $). 



\subsection{Other examples of the uses of $ D_S $.}

$D_S$ used to analyze venoms was published (discussed below, \cite{DSuze2010a, DSuze2015a}), although perhaps the most important factor was that when Complexity International, where our work \cite{Sevcik1998a} was originally published, ceased to be active and the work was deposited in www.arXive.org, \cite{Sevcik2010}. With the use of www.arXiv.org, GoogleSholar starts tracking article usage (see \url{https://scholar.google.com/citations?hl=en&user=0VhfpTIAAAAJ)}, but there are $D_S$ uses that are unreported in GoogleScholar. We are aware of multiple attempts to calculate $ D_{HB} $ and $ D_K $ before 2010, some of which are discussed here \cite{Gnitecki2004, Kantelhardt2008, Gneiting2012}
\begin{figure}[h!]
	\centering
	\includegraphics[width=8cm]{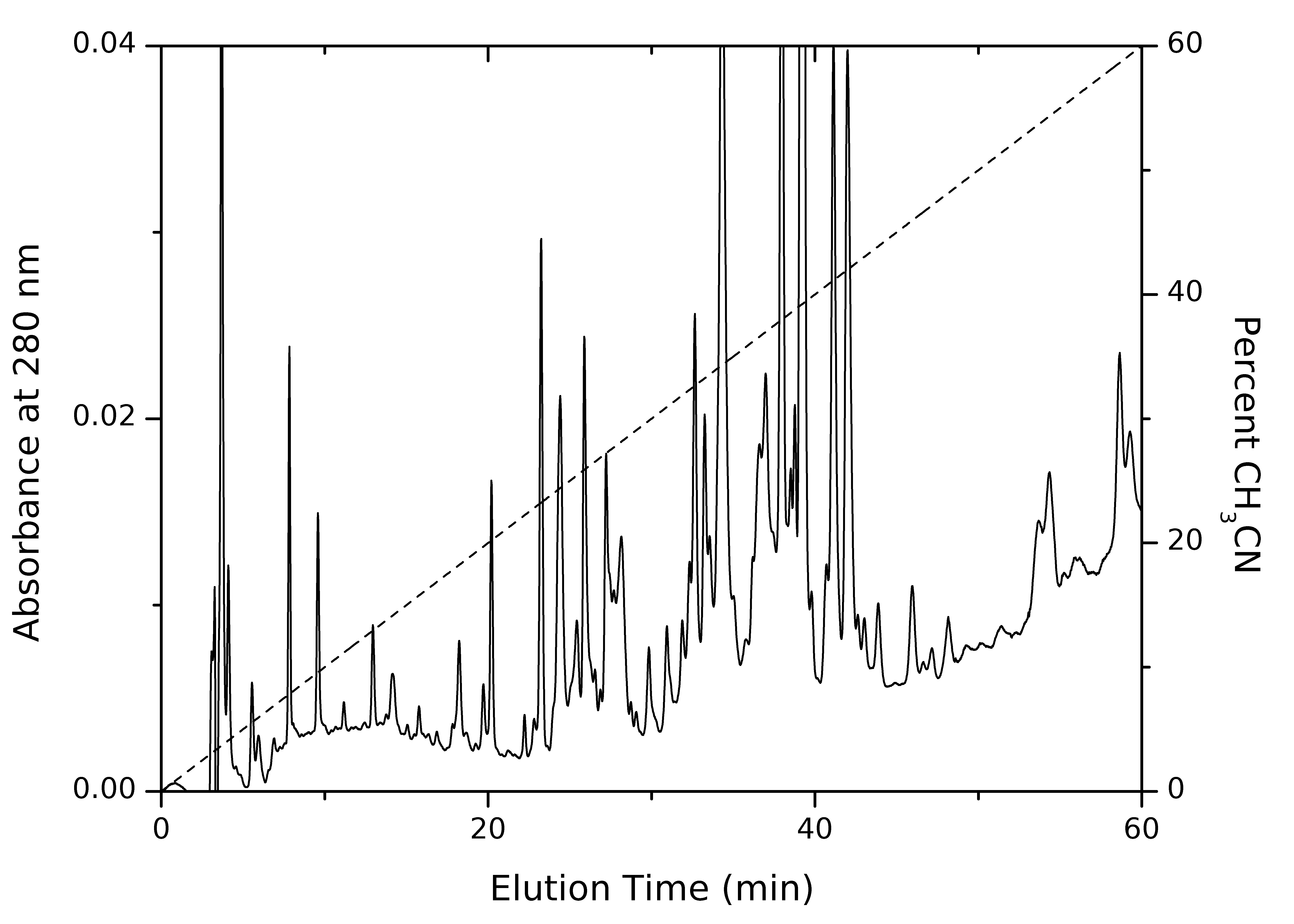}
	\caption{
		\footnotesize{
			\textbf{Reverse phase high-performance liquid chromatography (HPLC) of \textit{Tityus discrepans} scorpion venom.} HPLC separation of the water-soluble venom \textit{T. discrepans}  was made with a column of C\textsubscript{18} of reverse phase, the data were acquired at a frequency of 1 Hz, so that 60 min represent a $ n = 3600 $ points. The cknematographic profile shown here corresponds to a batch milking of individual venoms, of 100 specimens of \textit{T. discrepans}, put together as a lot poisons. For more details refer to the text of the book or the original publication \cite{DSuze2010a}.
		}
	}\label{F:ElucionTdiscrepans}
\end{figure}

\subsubsection{Using $ D_S$ to compare complicated systems.}

There are several situations where we consider complicated\footnote{The word \comillas{complicated} is used here to re preserve \comillas{complex} for numbers of type $(a + b\sqrt{-1}) \in \Com$.} systems that we need to compare. An example appears when we need to compare the components that are separated in various chemical methods that are usually grouped under the name of chromatography. In these methods, a set of very diverse molecules are diluted in a liquid or gaseous medium that is forced to move through a solid medium, with which they interact and by that interaction, they separate forming \comillas{peaks} that are collected to study their properties. 

One of the chromatographic techniques with the greatest capacity to separate compounds is the so-called high-performance liquid chromatography (HPLC), which separates components with different properties of electric charge, molecular weight or polarity \cite{Snyder1979}. There is a huge number of examples of this analysis, here we will limit ourselves to consider the example of its use for the analysis of natural poisons (called \scomillas{toxins}), produced by a genus of scorpions from South America, which we will arbitrarily limit to a genus called \textit{Tityus} of which only in Venezuela more than 50 species \cite{GonzalezSponga2001} with great toxicity to humans are known. 

\begin{figure}[h!]
	\centering
	\includegraphics[width=8cm]{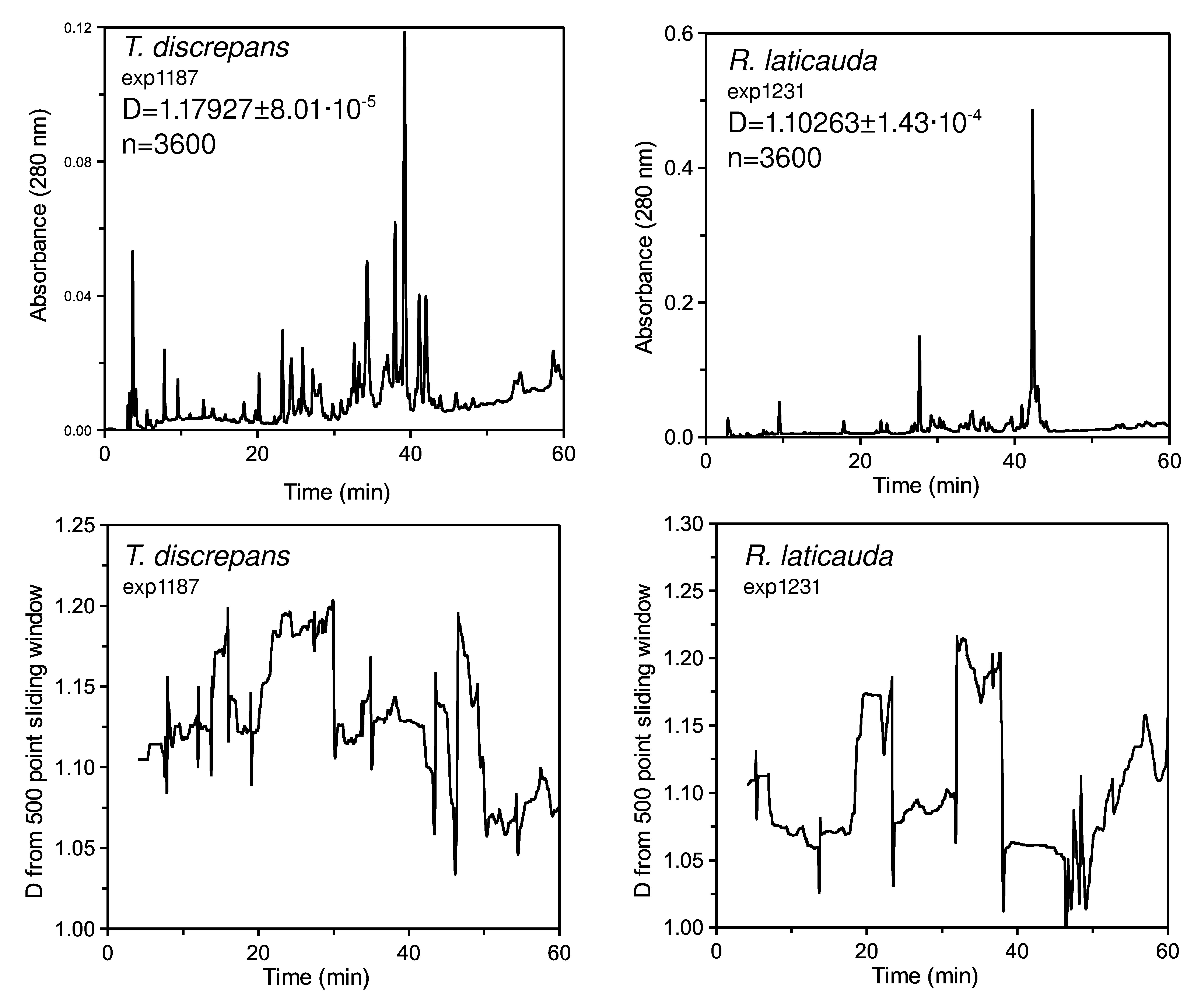}
	\caption{
		\footnotesize{
			\textbf{Two sources for calculating the fractal dimension ($ {D_S} $) of the elution pattern .} The figure presents data on the venoms of the scorpions \textit{Tityus discrepans} (panels on the left) and \textit{Rhopalurus laticauda} (panels on the right). The chromatography protocol is the same as that used in Figure \ref{F:ElucionTdiscrepans}. The top panel presents the pattern of elution an extraction (milking) of each venom of an animal of each species as ordered. Data were acquired at a frequency of $1 Hz$, which is equivalent to n=3600. The bottom panel presents the fractal dimension $ D_S $ calculated for the upper traces and $ D_S $ was $ 1.17927\pm 8.01\times10^{-5} $ [mean calculated with Eq. (\ref{E:D_S}) and standard deviation as $\sqrt{\var({D_S}})$ calculated with Eq. (\ref{E:VarDelta_y_LN}) for $ n = 3600 $ points of \textit{T. discrepans},and under the same conditions was $ 1.102637 \pm 1.43 \times 10^{-4} $, For \textit{R. laticauda}. If you use a sliding window of $ n $ points it is possible to calculate $ Q= D_s-1 $, assign that value to the center of the window and move it by 1 point, repeat the calculation of $ Q $ and assign it to the next point in the abscissa, and so on until the available points of the waveform are exhausted. This is what is presented calculated with a sliding window of $ n = 500 $ digitized points as explained for the upper panels, the value of $ D_S $ obtained from the first $ 500 $ points is assigned to the point $ x = 250 $, the window is offset by 1 point, it is recalculated and the value is assigned to the point with $ x = 251 $, and so on to the point $ N-250 $. For more details refer to Figure \ref{F:ElucionTdiscrepans}, the text of the article or the original publication \cite{DSuze2010a}.
		}
	}\label{F:HPLCyFractalDS}
\end{figure}

Here we will limit ourselves to consider the case of more impact, \textit{Tityus discrepans} that coexists with the largest city in the country, Carcas of 4 million inhabitants with about 5000 annual scorpion sting cases registered \cite{DSuze2010a, DSuze2012, DSuze2015, DSuze2015a}, but which is only one of the localities affected by scorpion ism by \textit{Tityus} in Venezuela. From the venom of \textit{T. discrepans} alone, some 206 fractions of toxic peptides \cite{Batista2006} have been separated. The problem of the complexity of these poisons is highly relevant. To separate this large number of compounds, it is usually required to rechromatograph (repeat chromatography) under modified conditions the compounds obtained under the initial conditions.

\textbf{Figure \ref{F:ElucionTdiscrepans}} presents the initial result of a chromatography of a batch composed of the milking of venom of 100 specimens of \textit{T. discrepans}. A large number of chromatographic peaks are observed, and different peaks often have different effects, peaks that appear together in the chromatography profile usually have similar effects \cite{DSuze2015}. With elution profiles of such complexity, comparing poisons is complicated. And the complexity increases if one considers that the chromatographic profile varies between individuals of the same species, and varies seasonally as well. 

\textbf{Figure \ref{F:HPLCyFractalDS}} presents the elution patterns of an individual (meang 1 milking from 1 scorpion)) milking of \textit{T. discrepans} venom and an individual milking of venom of \textit{R. laticuauda}, a species of scorpion common in areas of Venezuela below 500 meters above sea level, which poses no risk to humans. The figure is an example that scorpions can be compared from individual to individual with HPLC, so an instrument for comparing individual chromatography is very useful. The upper panels of Figure \ref{F:HPLCyFractalDS} show the chromatograms, \textit{T. discrepans}, above and on the left, and \textit{R. laticauda}, at the top right. Below each chromatogram is a graph of $D_S $ calculated with a sliding window of 500 points in length, as indicated in the text of the figure.

\begin{figure}[h!]
	\centering
	\includegraphics[width=10cm]{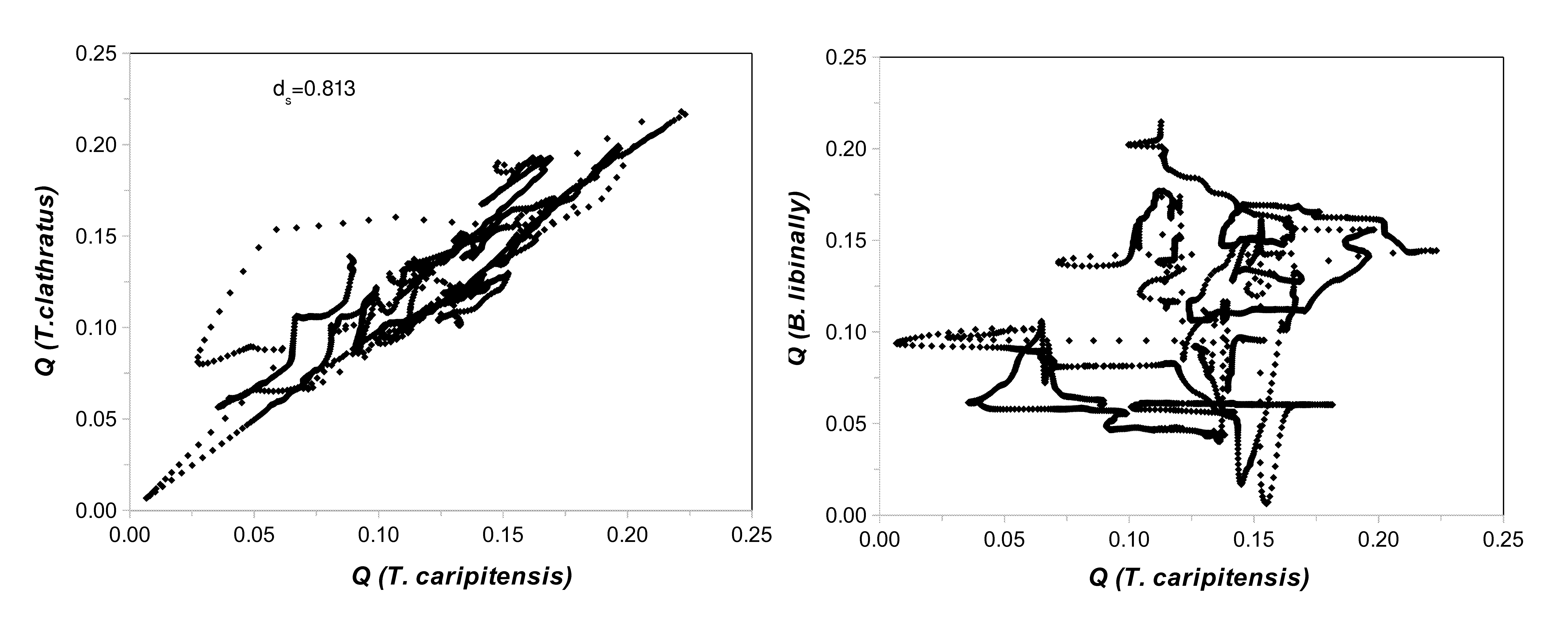}
	\caption{
		\footnotesize{
			\textbf{Tortuosity graph constructed with a sliding window $ \ngr{Q= D_s-1} $.} In cases such as the elution profiles of poisons, the waveform to be analyzed is obviously not infinite. In such cases fractal dimension can be questioned, and it is preferable to use the {\scomillas{tortuosity}} component of the fractal dimension that exceeds that of a line. If a sliding sliding window of $ n=500 $ is used as explained for Figure \ref{F:HPLCyFractalDS}, but here the set of $ \{Q_{1,i}\} $ is plotted against the set $ \{Q_{2,i}\} $, where $ 1 $ expresses one kind of scorpion and $ 2 $ the other; If both species were identical, the graph would be a line with slope $ 1 $ and intercept $ 0 $. The figure shows that \textit{T. caripitensis} and \textit{T. clathratos} have similar venoms, but that the venom of \textit{Broteas libinally} is very different from that of \textit{T. clathratus}). For more details refer to Figure \ref{F:HPLCyFractalDS}, the text of the book or the original publication \cite{DSuze2010a}.
		}
	}\label{F:Ch20UsoDeQ}
\end{figure}

\textbf{Figure \ref{F:Ch20UsoDeQ}} is an example of using $ Q $ curves, such as those shown in the lower panels of the \textbf{Figure \ref{F:HPLCyFractalDS}}. The figures were prepared by pairing point by point $ \{Q_{i,1}\}_{i=1,\ldots,N} $ with the points of $ \{Q_{i,2}\}_{i=1,\ldots,N} $ as ordinate. Which of the two sets was the  abscissa and which ordinate, were arbitrary decisions of the experimenter. It is easy to conceive that if $ \{Q_{i,1}\}_{i=1,\ldots,N} $ and $ \{Q_{i,2}\}_{i=1,\ldots,N} $ were equal, the points on the graph would form a diagonal with slope 1. In the left panel of the figure, the sequence for \textit{T. clathratus}  was used as ordered, and \textit{T. caripitensis} as abscissa. Although the correlation between the venoms is not perfect, the points are grouped around the line with slope 1 and intercept 0, is good enough to produce a coefficient of determination of 0.813. The graph on the left also has \textit{T. caripitensis}  as abscissa, but the ordinate corresponds to the venom of \textit{Brotheas libinalli} $ D_{S}$. The distribution of the points in the graph is far from its diagonal. The venom of \textit{B. libinalli} offers no danger to humans. The graph suggests that the venoms of \textit{T. clathratus} and \textit{T. caripitensis} are very similar, and that the venoms of \textit{B. libinalli} and \textit{T. caripitensis} are very different. The reader interested in more detail should refer to the urinals studies \cite{DSuze2010a, DSuze2015a}.

\subsubsection{Fractal dimension of service queues.}

There is an indefinite number of systems that provide services to an indefinite number of service seekers. Indefinite, here, has the same implication of $ \aleph_0 $, an \textbf{\textit{indefinitely large}} number fulfilling $\aleph_0 \in \Z$. Some services (the bus that passes at a certain time through my stop, the emergency of a hospital, the supermarket or pharmacy where I buy, etc.) are very obviously service provided to those demanding them. In all of them there is an obvious service provider, and a queue or line, of those waiting to be served. Perhaps less obvious as a service are public traffic means such as the, so-called highways. However, in all these cases the flow of those served follows a common function: a random Brownian march \cite{Brown1828, Brown1946, Samuel2012, Ossenbruggen2019, Muroki2019}.

Four years ago I was consulted at the La FuenfríaMadrid Hospital, a hospital of the Madrid Health Service (SERMANS) dedicated to treating chronic cases. The reason for the consultation was to increase the efficiency of the Hospital, prone to prolonged periods of low occupancy. Full data was published \cite{RodriguezHernandez2022} and we refer interested readers to that publication for details omitted here. For the analysis, daily Hospital occupancy data from May 1, 2014 to December 19, 2017 \cite[corrections for asymmeter errors as indicated here]{RodriguezHernandez2022}.

\begin{figure}[h!]
	\centering	\includegraphics[width=9cm]{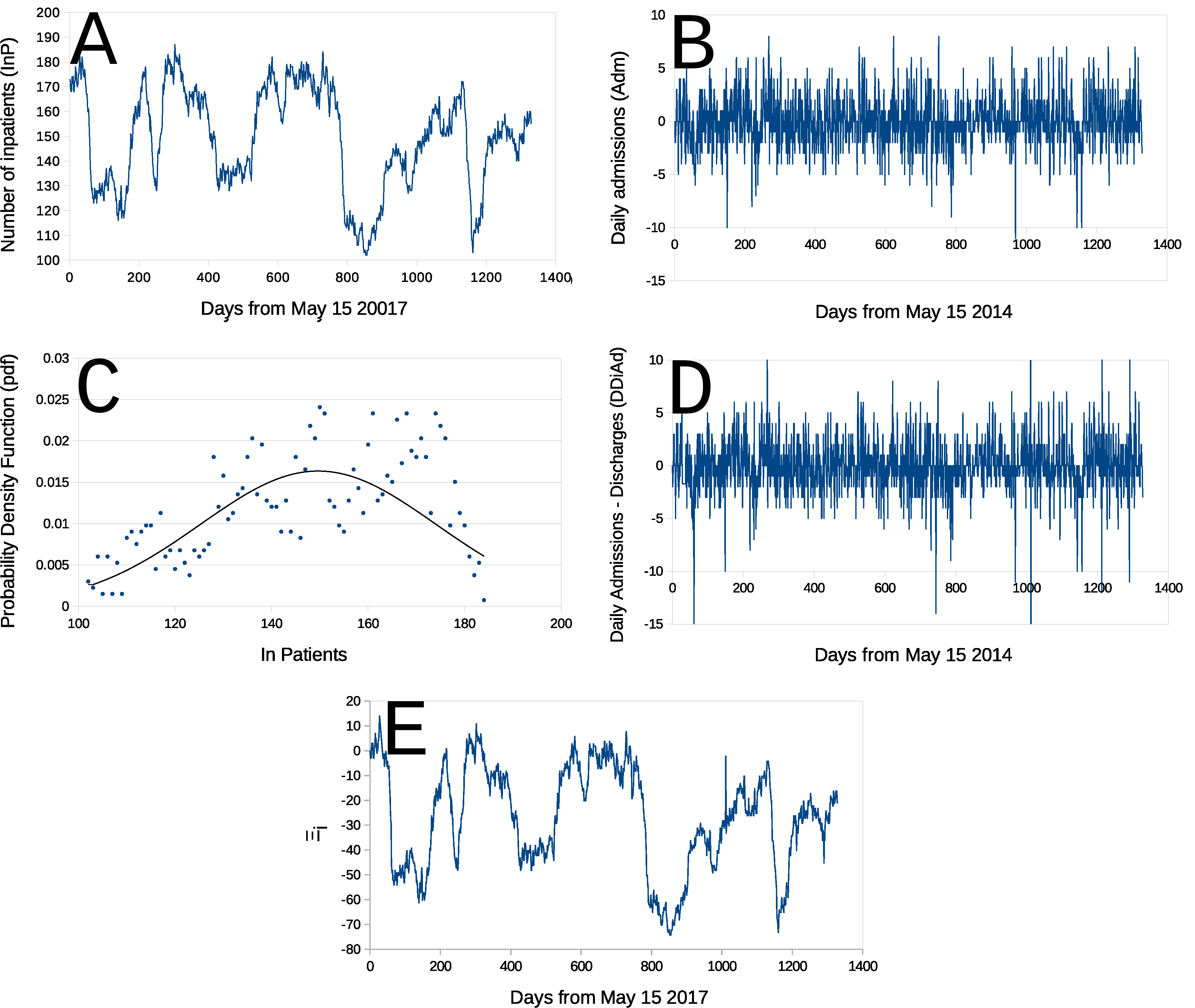}
	\caption{
		\footnotesize{
			\textbf{Patients hospitalized every day (InP) at La Fuenfría Hospital from May 1, 2014 to December 19, 2017.} The panels are: \textbf{Panel A}- Number of hospitalized patients (\textbf{InP}) at Hospital La Fuenfría on each day of the Hospital study; \textbf{Panel B}- Daily variations (\textbf{DInP}) of patients staying at Hospital [$ \Delta and $ in Eq. (\ref{E:SecDifGen})]; \textbf{Panel C}- Panel A data expressed as \textit{f}umtion of \textit{d}ensity of \textbf{p}robability (\textit{fdp}), the probability of observing a measurable or countable random event) for any number of patients (\textbf{InP}) who remain hospitalized each day during the study period ($ \bullet $) \cite{Wilks1962}, a Gaussian line of type ($ N[\bar{y}, s[y] $) with mean $ \bar{y} $ and variance ($ s[y]^2 $) both calculated for the 1329 data in Panel A ($ \bar{y}=149.2 $ and $ s[y]=24.4 $); \textbf{Panel D}- Daily differences between admissions (\textbf{Figure \ref{F:Admissions_Discharges}A}) and hospital discharges (\textbf{Figure \ref{F:DailyHospitalised}C}) at the Hospital (\textbf{DDiAd}). \textbf{Panel E}- Random march $ \ngr{\Xi_i} $ built for \textbf{DDiAd} (Panel D) using Eq. (\ref{E:DDiAd}), please note that the end resembles the \textbf{InP} sequence in Panel A. In all cases abscissa is the number of days between May 1, 2014. The largest negative or negative peaks are off-scale from the charts.
		}
	}\label{F:DailyHospitalised}
\end{figure}

With the Hospital data, a sequence of values of $ \Delta y_i $ was constructed as indicated in the following equation:
\begin{equation}\label{E:SecDifGen}
	\Delta y_{t}   =   \begin{cases}
		0 &\implies t=0\\
		y_{t} - y_{t-\Delta t} &\implies t > 0
	\end{cases}
\end{equation}
Here the $ y_i $ are daily data and the $ \Delta y_i $ are the daily changes, of the various occupations of patients in the Hospital. The \textbf{Figure \ref{F:DailyHospitalised}A} shows the sequence $ \Delta y_{t} $ calculated with the Eq. (\ref{E:SecDifGen}), the curve covers a total of 1216 days of study. During the study period, the number of hospitalized patients was $150\;(149-151)$ (median, and a 95\% CI, $n = $1329 days) in a range of ($102-187$) patients, meaning that Hospital occupancy is $0.77\;(0.53-0.77)$ (median, and, its 95\% CI). There could be one 4 \comillas{cycles} where the curve goes from 60\% from occupancy to a total of 98\% occupancy in a \comillas{cycles} of about 304 days ($ \approx 10 $ months) duration?. If we accept those \comillas{cycles} as real, they do not follow any annual seasonal period that we could identify. something that is, at least, rare.

\textbf{Figure (\ref{F:DailyHospitalised})C} is a graph of the pdf of daily variations of the data in \textbf{Figure (\ref{F:DailyHospitalised})A}, there it is observed that the data are not Gaussian, there are no negative data and there is a skewness towards positive high values. 

Trying to understand \textbf{Figure \ref{F:DailyHospitalised}A} we built \textbf{Figures \ref{F:DailyHospitalised}B} and \textbf{\ref{F:DailyHospitalised}D}. The first of these (\textbf{Figure \ref{F:DailyHospitalised}B}) is the number of patients the Hospital receives daily, and the second (Figure \ref{F:DailyHospitalised}D) is the number of patients the Hospital discharges (patients leaving the Hospital) daily. From the simple observation of the data in\textbf{Figures \ref{F:DailyHospitalised}B} and \textbf{\ref{F:DailyHospitalised}D}, there does not seem to be a pattern in them, they look like two \comillas{white noises} \cite{Hastings1993}, random oscillations around a mean, which are commonly distributed in Gaussian form.

\begin{figure}[h!]
	\centering	\includegraphics[width=9cm]{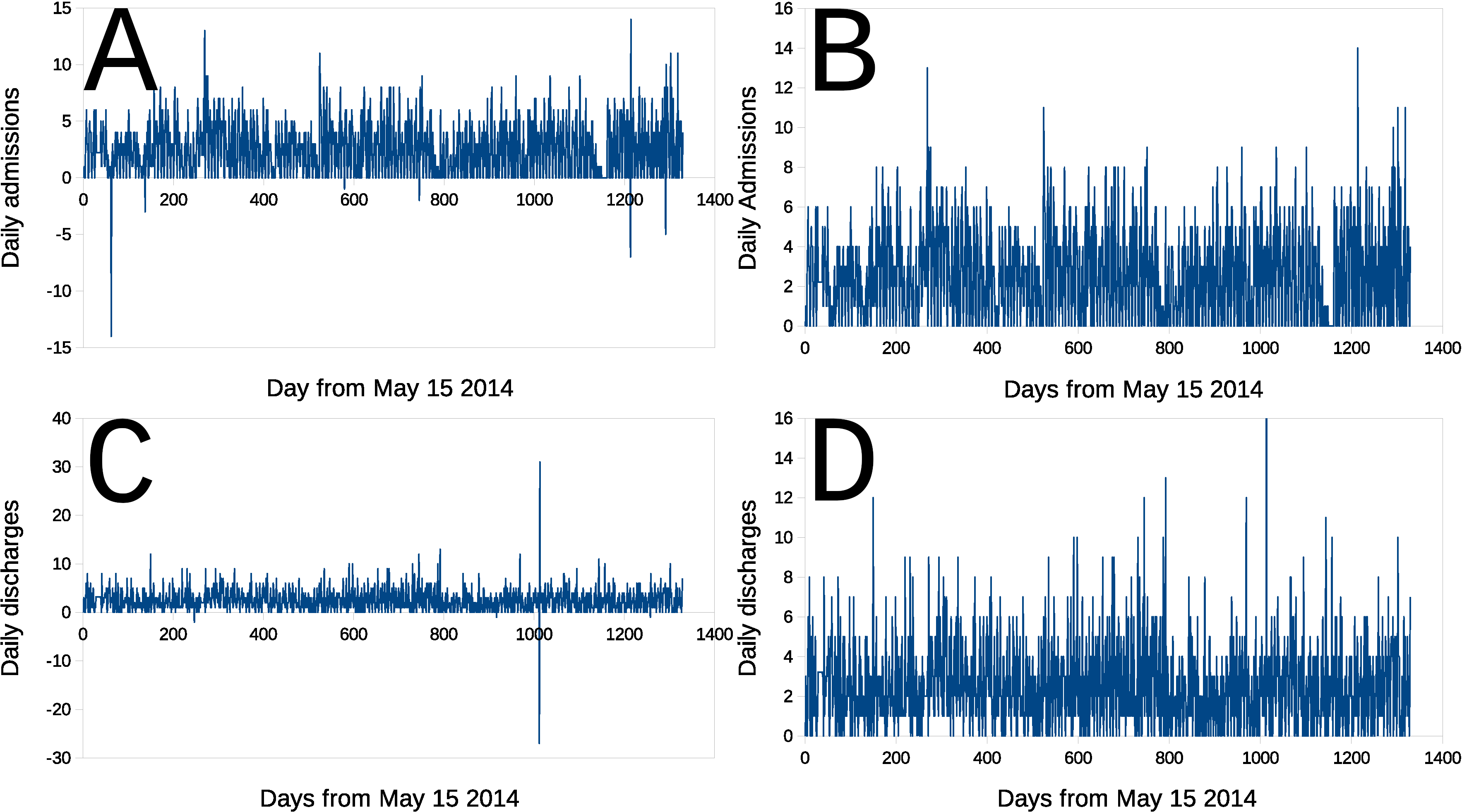}
	\caption{
		\footnotesize{
			\textbf{Sequences of admissions and daily discharges from La Fuenfría Hospital.} The Panels are: \textbf{Panel A}- Hospital Daily Admissions Sequences (\textbf{Adm}); \textbf{Panel B}- Same as panel A, but changing the scale to display data $ \geqslant0 $; \textbf{Panel C}- Sequence of daily discharges (\textbf{Dis}); \textbf{Panel D}- Same as Panel C, but scaled to include only data $ \geqslant0 $. In all panels, abscissa is the days elapsed since May 1, 2014, and the ordered abscissa is the daily number of patients admitted or discharged. Negative spikes represent transcription errors of data that were available for study, positive extreme data probably have the same, see original reference for more data  \cite{RodriguezHernandez2022}.
		}
	}\label{F:Admissions_Discharges}
\end{figure}

It is surprising, however, that if we calculate the daily difference between input and output and graph them, we obtain the \textbf{Figure \ref{F:DailyHospitalised}E}. This figure is constructed as the $ \Xi_i $ of the \textbf{Figure \ref{F:DailyHospitalised}E}, Built with Eq. (\ref{E:DDiAd}):
\begin{equation}\label{E:DDiAd}
	\ngr{ \Xi_i} = 
	\begin{cases} 
		0  &\implies i=1 \\
		DD_iAd_i + \ngr{ \Xi_{i--1}}&\implies 2 \leqslant i  \leqslant 1328 \; .
	\end{cases}
\end{equation} 
Surprisingly, with the exception of a few peaks that go out of the curve, probably resulting from the subtractive cancellation of small differences, the \textbf{Figure \ref{F:DailyHospitalised}E} is almost identical to the \textbf{Figure \ref{F:DailyHospitalised}A}. The fundamental difference between the curves in the \textbf{Figures \ref{F:DailyHospitalised}B} and \textbf{D} is that they reflect a delay in admitting a patient to the Hospital after another has been discharged, this shows the original work of Rodríguez-Hernández and Sevcik \cite[Check Figure 4]{RodriguezHernandez2022}. The Figures \textbf{\ref{F:DailyHospitalised}E} and \textbf{\ref{F:DailyHospitalised}A} are practically identical, and belie the idea that the apparent \comillas{cycles} are due to some seasonal factor external to La Fuenfría Hospital or the SERNAS hospital network of which the Hospital is a part. And more, the data favor the hypothesis that the oscillations of the occupation of the Hospital La Fuenfría, is due to factors that determine a delay in admitting a patient when another is discharged, not to some hidden factor external to the Hospital, or in the best of cases to the hospital network of SERMAS.

In \textbf{Figure \ref{F:Admissions_Discharges}} it becomes more evident that income and discharges have a positive median (\textbf{Figure \ref{F:Admissions_Discharges}B and D}) when a few negative data likely resulting from subtractive cancellation between small data are deleted (\textbf{Figures \ref{F:Admissions_Discharges}A and \ref{F:Admissions_Discharges}C}).

\begin{figure}[h!]
	\centering	
	\includegraphics[width=9cm]{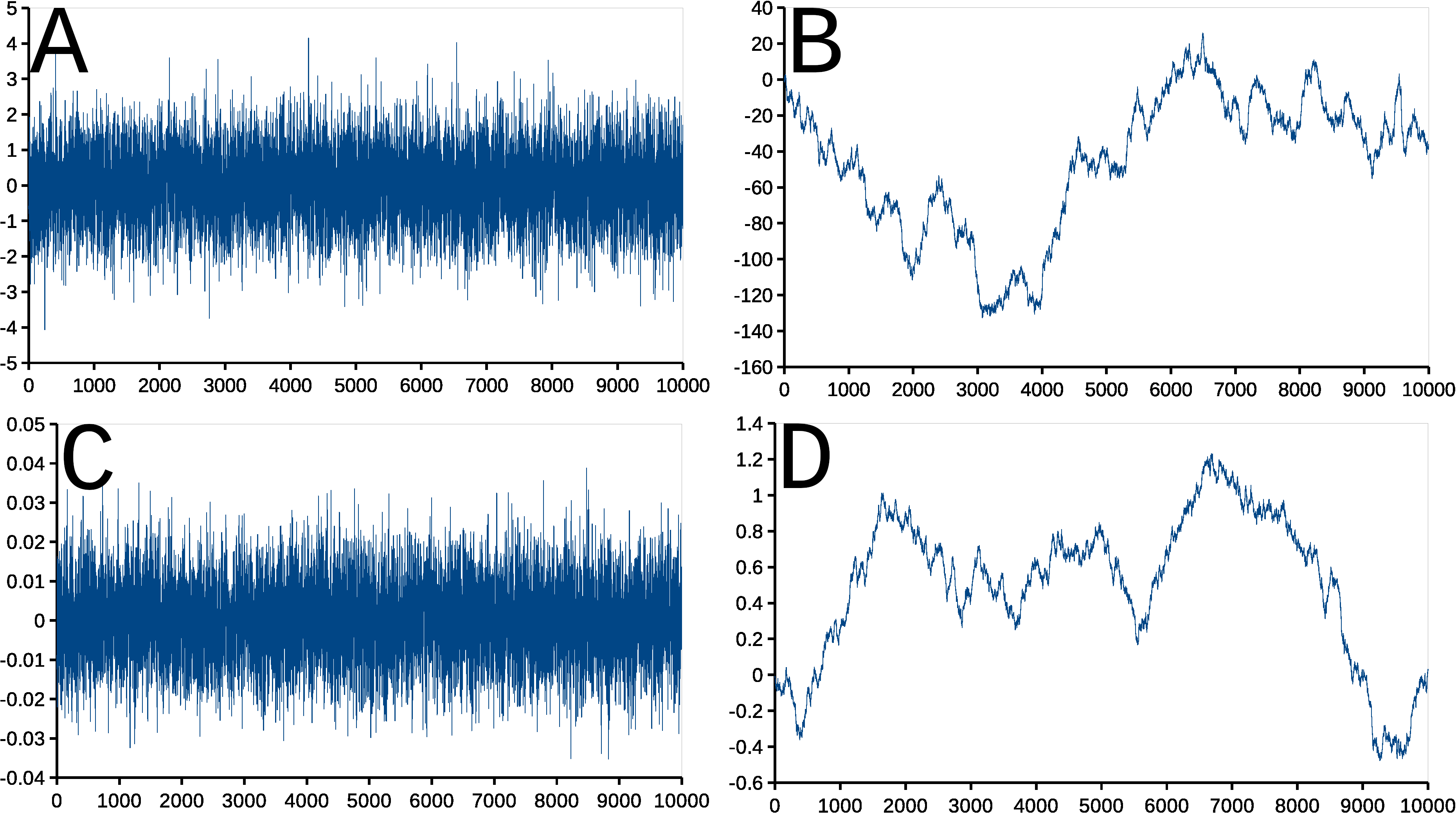}
	\caption{
		\footnotesize{
			\textbf{Two white boise sequences generated with Monte Carlo simulations [\textbf{Panels A} and \textbf{C}], as indicated by Box and Muller \cite{Box1958, Press2007}}. The Panels are: \textbf{Panel A}, {Gaussian White Noise} and \textbf{B}: {Gaussian Brownian Noise \textit{Rw}}, generated using Eq. (\ref{E:BroGaus}) forms the data in panel A;\textbf{Panel C}, {Gaussian white noise} and $ \{g_i = \G{0,10^{-4}}\}_{i=1,2, \ldots,10^4} $; \textbf{D}: {\textit{Rw}} Gaussian, generated using the Eq. (\ref{E:BroGaus}) with panel C data, the abscissa and ordinate are arbitrary. Please note that $\left[D_S \pm s(D_S) \right]$ estimated with Eqs. (\ref{E:D_S}) and (\ref{E:VarDelta_y_LN}) are: Panel A, $ 1.66122 \pm 7.60191 \cdot 10^{-4} $; Panel B, $1.32692 \pm 7.6926\cdot 10^{-4} $; Panel C, $ 1.6692 \pm 7.64176\cdot 10^{-4} $; Panel D, $ 1.31723 \pm 7.72806\cdot 10^{-4} $. \textit{Rw} is a \textit{random walk}, of the original work. See the original reference and text of this book for more data and details \cite{RodriguezHernandez2022}
		}
	}\label{F:Gauss} 
\end{figure}

The relationship between \textbf{Figures \ref{F:DailyHospitalised}A} and \textbf{\ref{F:DailyHospitalised}E} can be understood if one uses {Monte Carlo simulation} to generate data around a known pdf and {Marcovian time series} \cite{Meyn1993, Shi2012}. In principle, we can visualize two kinds of time series where data that have a pdf appear such as $ y_i = g(x_i \given  \{\theta\}) $, where $ \{\theta\} $ is a set of parameters on which $ y_i $ depends. Then we can conceive of a series such as:
\begin{eqnarray}
	y_i &=& g(x_i \given \{\theta\}) \label{E:SomeWhiteNoise} \\
	y_i &=& y_{i-1} + g(x_i \given \{\theta\}) \label{E:SomeMarkowNoise}.
\end{eqnarray} 
In the first case any point $ (y_j \lindep y_i)_{\forall (i \neq j)} $\footnote{$\forall$ means \scomillas{for all}}, is a waveform we call {white noise}. In the second case $ (y_i \nlindep y_j)_{\forall (i \neq j)} $, but we cannot say how, this is what is called a {Markovian process} or a {Markov series}, the sequence that constitutes the waveform is usually called Brownian noise, since that kind of movement was discovered in the nineteenth century as characteristic of pollen grains in liquid medium by Brown \cite{Brown1828, Brown1829}. In white and Brown noises of the most common type, however, the pdf is Gaussian such that $ g(x_i \given \{\theta\}) = \G{\mu, \sigma^2} $, with mean $ \mu $ and variance $ \sigma^2 $. Brownian noise follows a function as:
\begin{equation}\label{E:BroGaus}
	b_i = \begin{cases}
		b_1 = 0\\
		b_{i} + g_{i-1} \implies 2 \geqslant i \geqslant 1329 .
	\end{cases}
\end{equation}

Please note that \textbf{Figures \ref{F:Gauss}A} and \textbf{\ref{F:Gauss}B} were calculated with Eq. (\ref{E:BroGaus}) are plotted with \textbf{exactly the same set $ \ngr{\{g(x_i \given \{\theta\})\}} $} obtained from the Monte Carlo simulation, just as \textbf{Figures \ref{F:Gauss}C} and \textbf{\ref{F:Gauss}D} do with yours. 

This type of noise in its white and Brownian version, also referred to as Brown noise, is presented in the \textbf{Figure \ref{F:Gauss}}. \textbf{Figures \ref{F:Gauss}A} and \textbf{\ref{F:Gauss}B}, are white and Brown noise with a pdf $ \G{0.1} $ and the \textbf {Figures \ref{F:Gauss}C and \textbf{\ref{F:Gauss}} with $ \G{0.10^{-4}} $}. Note that if you look at the \textbf{Figures \ref{F:Gauss}A} and \textbf{\ref{F:Gauss}C}, or the \textbf{Figures \ref{F:Gauss}B} and \textbf{\ref{F:Gauss}D} without noticing the ordinate scale or variances noted above, one of those pairs looks the same as each other; the same goes for \textbf{Figures \ref{F:Gauss}B} and \textbf{\ref{F:Gauss}C}; These similarities are the so-called {self-similarities} or {self affinities} of fractal processes, which in the case of white and Brown noises determine that they look the same at any scale of the abscissa that are observed. Note that despite the differences in their $ g(x_i \given \{\theta\}) $ figures \textbf{\ref{F:Gauss}B} and \textbf{\ref{F:Gauss}D} look very similar to each other and to \textbf{Figures \ref{F:DailyHospitalised}A} and \textbf{\ref{F:DailyHospitalised}E}.

The fractal dimension $ D_{HB} = 2 $, for a white noise and is $ D_{HB} = 1.5 $ for one Brown noise \cite{Hastings1993, Sevcik1998a, Sevcik2010, Sevcik2017b}. Rodriguez-Hernandez and Sevcik \cite{RodriguezHernandez2022} showed that $ D_S \rightarrow 2 $ for \textbf{Figures \ref{F:Gauss}A} and \textbf{\ref{F:Gauss}C} when the sequences lengthen towards $ \aleph_0 $ and that at the same time $ D_S \rightarrow 1.5 $ when the noise sequences \textbf{Figures \ref{F:Gauss}B} and \textbf{\ref{F:Gauss}D} are also lengthened. Rodríguez-Hernández and Sevcik \cite{RodriguezHernandez2022} is therefore a demonstration of the existence, perhaps unsuspected, that in the operation of a modern hospital, there may be hidden factors, which make it behave like most service providers as a queue in contemporary literature in English), which in this case determines periods of low use of the Hospital, which can be considered \comillas{inefficiency} but which are not soluble if the hospital network of Madrid en bloc is not considered, not as individual hospitals of SERNAS. Please review the original work \cite{RodriguezHernandez2022} for more details.
\begin{figure}
	\centering	
	\includegraphics[width=5cm]{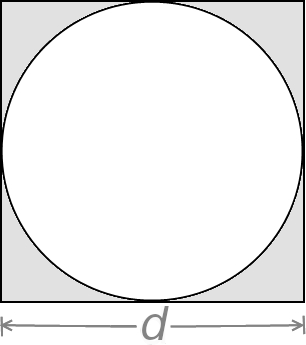}
	\caption{
		\footnotesize{
			\textbf{The relationship between open boxes and open circles to calculate the fractal dimension of a curve.} If we call $ \epsilon $ to the radius of the circle, then the side of the square will be $ d=2 \cdot \epsilon $. Please note that the lines have been drawn very thickly, they should be points of radius $ \epsilon = \frac{d}{2} \rightarrow 0 $ and the outermost layer of points, in an \comillas{open} square or a circle, are not included in the respective square or a circle. From the figure it is intuitively obvious that, except for possible segments of a curve parallel to the abscissa or the ordinate of the curve, it will take fewer squares than circles to cover the curve since $ d^2 > \pi \epsilon^2 = \pi \left( \tfrac{d}{2}\right) ^2 \impliedby 1 >  \tfrac{\pi}{4} $. Where $ \epsilon $ is used as in Eq. (\ref{E:DimHausBes}) and subsequent ones.
		}
	}\label{F:CircleSquare} 
\end{figure}

\subsection{Higuchi\s fractal dimension.}

In the fractal literature \cite{Hastings1993} there are two ways to approximate the fractal dimension of a waveform. One of them is the use of open circles that cover the curve, this we have discussed in relation to $ D_S $ \cite{Sevcik1998a, Sevcik2010} where a double linear transformation to a unit square is used. The other is the use of \comillas{open boxes} that cover the wave, those boxes are usually squares of side $ d $. Long ignored by the authors, there is another form of fractal dimension estimation ussing boxes, developed by by Higuchi \cite{Higuchi1988, Higuchi1990}. It is called fractal dimension because the waveform whose fractal dimension is to be estimated is covered by \comillas{boxes}, actually, in two dimensions, \textit{squares }of side $ d $. As follows from \textbf{Figure \ref{F:CircleSquare}} it takes more circles than squares to cover any waveform other than a vertical or horizontal line. In other circumstances $ D_\blacksquare \lessapprox D_\bullet $. Here we use a slightly different notation than Higuchi \cite{Higuchi1988}, using matrix notation. To understand the fractal dimension of Higuchi \cite{Higuchi1988, Higuchi1990} we will consider a time series of observations such as 
\begin{equation}\label{E:Higuchi1}
	\{ x_1, x_2,\ldots, x_N\} .
\end{equation} 
From this sequence we will construct a subsequence such as
\small{
	\begin{equation}\label{E:Higuchi2}	
		\begin{matrix}
			\{X_k^m\} =& \left\lbrace  x(m) , x(m+k) , x(m+2k) , 
			\ldots , x\left( \cancel{m}+\left[ \frac{N-\cancel{m}}{\cancel{k}}\right] \cdot \cancel{k}\right)  \right\rbrace _{m=1,2,\ldots,k} \\
			\ngr{\then \{X_k^m\} =}& \ngr{\left\lbrace  x(m) , x(m+k) , x(m+2k), 
				\ldots , x\left(N\right)  \right\rbrace _{m=1,2,\ldots,k}} \quad \lqqd\\ 
		\end{matrix}
	\end{equation}
}
The deletions of the last term in Eq. (\ref{E:Higuchi2}) are mine. 

Here Higuchi introduces a condition that he does not explain, this is that the term in square brackets, there is none other than $ \left[ \tfrac{N-m}{k}\right] $, \comillas{denotes Gauss's notation} (there is no other explanation about Gauss suggesting that) and that both $ k $ and $ m $ are integers [we presume that it means $ (m \in \Z) \land (m \in \Z) $] and points out that $ m $ and $ k $ indicate the initial time and the time interval (Between the points?) respectively. Starting from that series Higuchi \cite{Higuchi1988} constructs the following matrix of equations, for say $ N = 100 $:
\begin{equation}
	\begin{matrix}
		\{X_1^3\}&=&\left\lbrace x(1),x(4), x(7), \ldots, x(97),x(100)\right\rbrace \\
		\{X_2^3\}&=&\left\lbrace x(2),x(5), x(8), \ldots, x(95),x(98)\right\rbrace \\
		\{X_3^3\}&=&\left\lbrace x(3),x(6), x(9), \ldots, x(96),x(99)\right\rbrace \\
	\end{matrix}
\end{equation}

Higuchi \cite{Higuchi1988} \textit{defines (without explanation)} the length of each curve $ \{X_k^m\} $ as follows:
\begin{eqnarray}
	L_n(k) &=& \tfrac{ \sum_{i=1}^{\tfrac{N-m}{k}} \abs{x(m+i\cdot k)-x(m -(i-1)\cdot k)} \cdot \left( \tfrac{N-1}{\left[\tfrac{N-m}{\cancel{k}} \right] \cdot \cancel{k} }\right) }{k} \label{E:Higuchi_Length}\\
	L_n(k)&=& \frac{N-1}{[N-m] \cdot k } \cdot \sum_{i=1}^{\frac{N-m}{k}} \abs{x(m+i\cdot k)-x[m -(i-1)\cdot k]}  \label{E:Higuchi_Length_Simp} 
\end{eqnarray} 
Eq. (\ref{E:Higuchi_Length}) in its first form is the Higuchi curve length equation with more contemporary notation, Eq. (\ref{E:Higuchi_Length_Simp}) is the same, simplified to make it more understandable and compact. If the original publication \cite{Higuchi1988} is reviewed, there is no explanation about the two vertical lines that limit the term within the sum of the Eqs. (\ref{E:Higuchi_Length}) or (\ref{E:Higuchi_Length_Simp}), the journal where the article was published had no problem representing parentheses, braces or brackets of any size, as is obvious in Higuchi's article \cite{Higuchi1988} itself, so that these unexplained vertical lines can only be considered as indicating that the summation is made over the \textit{absolute values} of the differences within the summation. This is strange, assuming $p_i$ and $p_{i+1}$ are point in a two dimensional Euclidean space with coordinates $(x_i, y_i)$ and $(x_{i+1}, y_{i+1})$, respectively. The distance between $p_i$ and $p_{i+1}$, according to the Pitagoras theorem will be $\sqrt{(x_{i+1}- x_i)^2 + (y_{i+1}- y_i)^2}$ and the absolute value function is never necessary.

According to Higuchi \cite{Higuchi1988}, the fraction $\frac{N-1}{[N-m] \cdot k } $ represents the \comillas{\textit{normalization of curve length}} of a \comillas{\textit{time series subsystem}}. Also, according to Higuchi \cite{Higuchi1988}: the length of the curve is defined for the time interval \comillas{$ k$, $ \; \left\langle L (k) \right\rangle $}, as the average value of the parameter $ k $ of the sets $ L_m(k) $. Again according to Higuchi \cite{Higuchi1988}: if
\begin{equation}\label{E:RelExpD}
	\left\langle L(k)\right\rangle \propto k^{-D_{Hig}} , 
\end{equation} 
the {fractal dimension} of the curve is $ D_{Hig} $\footnote{The subscript \textit{Hig} is mine}. 

Thus the fractal dimension is a constant of proportionality between $ k $, the abscissa and $ L (k) $, the ordinate, the relationship between them is an exponential. The idea, however, is not original to Higuchi

Higuchi \cite{Higuchi1988} validated his method to determine the fractal dimension tests with \comillas{simulated data} to which he applied his method. First, he applies his technique to the simulated data $ \{y_i\}_{ i = 1, 2..... N} $ with fractal dimension $ D = 1.5 $, $ y_i $ is generated as
\begin{equation}\label{E:SimHiguchi}
	y_i = \sum_{j=1}^{1000+i} z_j
\end{equation}
where $ z_j $ is \textit{is assumed} Gaussian  with mean 0 and variance 1 \cite{Higuchi1988}. \comillas{The value of $1000 $ was taken arbitrarily to eliminate the effect of arbitrarily sampling $ z_i $ for $ y_1 $} \cite{Higuchi1988}. Higuchi used the following values for the interval $ k = 1,2,3,4 $, and {$ k = \left( 2^{\frac{j-1}{4}}\right)_{j= 11,12,13\ldots}  $ for $ k>4 $}, \comillas{where  Gauss notation} (\textit{sic}, I don't know what Higuchi means), here Higuchi is inconsistent now uses ($ k \in \R $), observe that $ k=2^{2.5} = 5.65685 \ldots $, for example. 

Then, if $ \left\langle L(k) \right\rangle $ is plotted against $ k $ on a logarithmic double scale, the data must fall on one on a line wcon slope $ - D_{Hig} $ (here we deviate from Higuchi, in the terms after the implication),
\begin{equation}\label{E:RectaHiguch}
	\begin{split}
		\log( L(k)) &= -D_{Hig} \log(k) \underset{\text{\tiny{con  suerte}}}{\implies} \\ D_{HB} &= \underset{d \rightarrow 0}{\lim} \; D_{Hig} = -\: \underset{d \rightarrow 0}{\lim}\left[  \frac{\log{L(d)}}{\log(d)}\right] = -\: \underset{d \rightarrow 0}{\lim} \; \log_d\left[{L(d)}\right] 
	\end{split}
\end{equation}
which is a form of Eq. (\ref{E:RelExpD}) changing the proportion to equality, which means that any difference produced by this change falls within Higuchi's $ D $. The $ d $ is entered with the same sense as it has in the \textbf{Figure \ref{F:CircleSquare}}. I introduce the implication because Higuchi simply calls \scomillas{fractal} his $ D $ \cite{Higuchi1988}, and it is Mandelbrot \cite[pg. 15 and Ch. 39]{Mandelbrot1983} who invents the term \comillas{fractal}, and associates it with the Hausdorff-Besicovitch dimension \cite{Hausdorff1918, Besicovitch1929}. The \comillas{luck} referred to indicates the lack of \textit{formal} association between $ D_{Hid} $ and $ D_{HB} $ that is not formally established anywhere by Higuchi. Higuchi \cite{Higuchi1988} simply \textit{defines}: $ D_{HB} = D_{Hig} $.

Higuchi \cite[pg. 280]{Higuchi1988} gives an example of his definition of length in relation to Burlaga and Klein\s description \cite{Burlaga1986} of these authors' fractality of the interplanetary magnetic field, as follows
\begin{equation}\label{E:BurlagaKleinHiguchi}
	\begin{split}
		L_{BK}(3) &= \tfrac{1}{9} [ 
		\abs{(x_4+x_5+x_6) - (x_1+x_2+x_3)} \\
		&\quad + \abs{(x_7 + x_8 + x_9) - (x_4 + x_5 + x_6)} \\
		&\quad + \abs{(x_{10} + x_{11} + x_{12}) - (x_7 + x_8 + x_9)} \\
		& \qquad \vdots \\
		& \quad + \abs{(x_{\aleph_0-2} + x_{\aleph_0-1} + x_{\aleph_0}) - (x_{\aleph_0-5} + x_{\aleph_0-4} + x_{\aleph_0-3})} ]
	\end{split}
\end{equation} 
which was here rearranged here to make it more compact and understandable. Note that in the last term the expression up to $ \aleph_0 $ is explicitly presented (to say a cardinal as large as you like). The original version of Higuchi only says \comillas{$ + \ldots \text{)} $} where a bracket seems to be missing. Then the same definition, with apparently the same errors, is repeated for $ \left\langle L(k) \right\rangle $. All definitions of Higuchi are strange. It is not stated, for example, why in $ L_{BK}(3) $ (Eq. \ref{E:BurlagaKleinHiguchi}) or $ \left\langle L(k) \right\rangle $ (Eq. \ref{E:Higuchi_Length_Simp}) the whole sum is divided by 3 ($\ldots/3$ at the end of each of those expressions) and each of the terms of the sum is divided by $ 3 $ as well, nor why division by this constant is not taken as a common factor of the sum to simplify it. In my transcription (Eq. \ref{E:BurlagaKleinHiguchi}) of the Higuchi equation I extracted them and placed them as $ \tfrac{1}{9} $ before the sum of the differences of the absolute values but I guess that $ \tfrac{1}{3} $ would be more correct.

Another incomprehensible thing is why Higuchi uses the absolute value of the differences between neighboring points to calculate the length $ L(k) $ of his curves. If these curves are constituted (away from Higuchi's notation) by a set of data pairs $ \{x_i,y_i\}_{i=1,2,\ldots,N} $ then the length of the curve will be 
\[ \mf{L}= \sum_{i=2}^{N} \sqrt{ (y_i - y_{i-1})^2 + (x_i - x_{i-1})²} \]
according to the Pythagorean theorem $L= \sum_{i=2}^{N}\abs{y_i - y_{i-1}}$, as Higuchi seems to use \cite{Higuchi1988}, so that $ \mf{L}>L $, which means that Higuchi \cite{Higuchi1988} underestimates the length of the curve\textquoteright{}s fractal dimension it claims to estimate, unless the author uses the two vertical bars as something distinct, and unexplained.

\subsubsection{A final word on Higuchi fractal dimension.}

The logarithms of waveform\textquoteright{s} length at a long time, $ \log_b\left\langle L (k)\right\rangle $ ($ b $ is any basis), for a time series, $ y_i $ with $ N = 217 $, is plotted as a function of $ \log_b L( k) $ in Figure 1 of Higuchi \cite{Higuchi1988}. This appears to be a kind of {Monte Carlo simulation} of a Brownian noise like the one shown and discussed in \textbf{Figure \ref{F:Gauss}B} and \textbf{ \ref{F:Gauss}D}, but does not give any information about the reliability of the generator of a random variable such as $ \G{0, \sigma^2} $ and today would not be considered acceptable for publication \cite{Sevcik1998a, Sevcik2010, Sevcik2016, RodriguezHernandez2022}. Higuchi \cite{Higuchi1990} makes a second publication with his method, in it he cleans up algebra somewhat and makes it less baroque. But again the 1990 \cite{Higuchi1990} publication he does what looks like a Monte Carlo simulations which is not described. They have been described for today's requirements. Monte Carlo, it seems. After the work of 1990 \cite{Higuchi1990}, the use of the \comillas{fractal dimension of Higuchi} disappears from the literature (in my experience) until the year 2004 when it reappears, and is  used used mainly in medicine, without any additional formal analysis \cite{Anier2004, Doyle2004, Klonowski2005, Spacic2011, Spasic2011a, Harne2014, Alnuaimi2017, Cukic2018, Liehr2019, Liehr2019a,  Shamsi2021, HiguchiDimension2022}.

Thus, ways of estimating $ D $ for curves in two-dimensional Euclidean spaces began to appear developed by Higuchi \cite{Higuchi1988, Higuchi1990} and by Katz \cite{Katz1988}. Both of Higuchi's papers were published in \textit{Physica D: Nonlinear Phenomena} \cite{Higuchi1988, Higuchi1990}, backed by another Japanese physicist.As we discussed earlier, the papers by Higuchi \cite{Higuchi1988, Higuchi1990} are strange. The first paper with a \comillas{muddy} algebra and many unexplained \textit{definitions}.
Higuchi's method, then, disappears from the literature for 15 years and is cited again after 2004 \cite{Anier2004, Klonowski2005, Harne2014, Alnuaimi2017, Cukic2018, Liehr2020, Shamsi2021, HiguchiDimension2022}, some of these references include attempts to validate Higuchi\s dimension \cite{Liehr2020}.

\section{Frequency analysis as a form of data analysis.}

Another interesting way for data analysis of time series, used for example  in Rodriguez-Hernandez and Sevcik \cite{RodriguezHernandez2022}, is frequency analysis. Frequency analysis is not part of the analysis of chaotic or fractal systems, and also includes strictly statistical time series concepts such as autocorrelations and autocovariances. We will continue to use the work \cite{RodriguezHernandez2022} as an example of this analysis.se can refer to \cite{Oppenheim1975, Bendat1985, Smith1997} among others.

\begin{figure}[h!]
	\centering	
	\includegraphics[width=8cm]{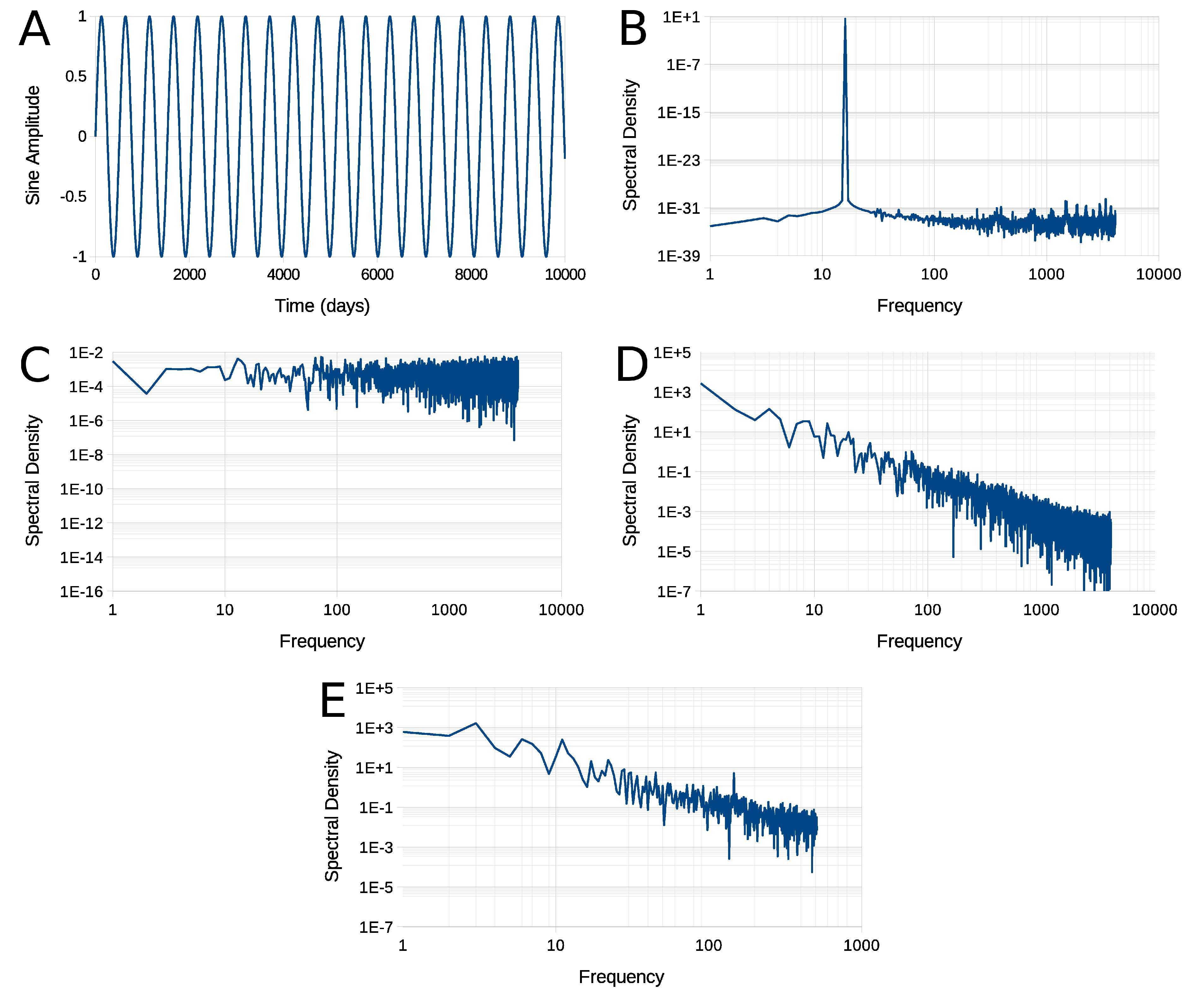}
	\caption{
		\footnotesize{
			\textbf{Analysis of the false periodicity of \textit{Rw}s. Using spectral density analysis ($ \ngr{ \psi_{f_j}} $).}  The panels are: \textbf{Panel A}- A perfectly periodic signal, the trigonometric function $ \zeta(x)= \sin(x \tfrac{ \pi }{128}) $; \textbf{Panel B}- $ \psi_{f_j} $ of the sine function displayed in panel A; \textbf{Panel C}- $ \psi_{f_j} $ of white Gaussian noise shown in Figure \ref{F:Gauss}A; \textbf{Panel D}- $ \psi_{f_j} $ of Brown's noise \cite{Hastings1993} \textit{Rw} shown in Figure \ref{F:Gauss}B; \textbf{Panel E}- $ \psi_{f_j} $ of the sequence of hospitalized patients ($ \ngr{ InP} $) that appears in the \textbf{Panels A} up to \textbf{E} of \textbf{Figure \ref{F:DailyHospitalised}A}. The ordinate of the panels from B to E, is the power dissipated at each frequency. The function shown is the displayed sine function presents the dissipated energy all its energy in a single peak $ \psi_{j} $ (Panel B). The \textit{sequence for the number of hospitalized patients $ SD{f_j} $ corresponds to a Gaussian \textit{Rw} without any underlying periodicity}. All graphics $ SD $ were fitted a window of Hann \cite{Blackman1959, Harris1978}. The rapid oscillations observed to the right from \textbf{Panel B to D} are artifacts due to the sampling of discrete points, in the signal to perform the Furrier transformation. \textit{Rw} is used in the original work cited as \textit{random walk}, more details in the text of this book and the original work cited  \cite{RodriguezHernandez2022}
		}
	}\label{F:SpectralDensities}
\end{figure}

A classic way to show periodicities of mathematical functions in the time domain is to transform them to the frequency domain, which is a sum of sines and cosines, using a fast Fourier transformation (FFT) \cite{Blackman1959, Welch1967}. The transformed values for each frequency ($ f $) are \comillas{complex numbers}, $c_f \in \Com $, with two components: one of them is some class sometimes called simply \comillas{real component} such as $ a_f \in \R $ and a $b_f \in \Im$ component called \comillas{imaginary}. A complex number has the form
\begin{equation}\label{E:CompNumb}
	c_f = a_f + b_f \sqrt{-1} = a_f + b_{f}  \J  \implies
	\begin{cases}
		c_f  &\in \{ \Com  \} \\
		a_f  \land b_f &\in \{\Re\} \\
		b_f\ngr{\iota} & \in  \{\Im\},
	\end{cases}
\end{equation}
where parameters indicate sets of numbers, and $ \in $ reads \comillas{\textit{belongs to}}. The \comillas{energy}, expressed as a stable nonzero curve, or as increases in sharp peaks or variations in amplitude (such as the peak in the \textbf{Figure \ref{F:SpectralDensities}B}, this is called \comillas{spectral density} ($\ngr{\psi_f}$) and, sometimes, we might informally call as the \comillas{weight}) , contributed by each frequency component to the energy of the total signal in a complicated process that fluctuates and is expressed as  
\begin{equation}\label{E:PowSpect}
	\left( \ngr{\psi_f} = \sqrt{a_f^2 + b_f^2}\right)  \in \R
\end{equation}
If you plot $ \ngr{\psi_f} $ against $ f $ you get a {a spectral density graph}, which is a set of real numbers ($ \{x_i\} \in \R $).
An intuitive way to understand this is to consider a perfectly periodic signal such as a sinusoidal trigonometric series such as a sine function $ \zeta (x) = \sin( \tfrac{2 x \pi}{T}) $ where $ \pi = 3.1415926 \ldots $ and $ T $ is a constant (called \textit{period}) that has the same units (meters, seconds, grams, volts, or any other) that has $ x $. An example is shown in \textbf{Figure \ref{F:SpectralDensities}A}. \textbf{Figure \ref{F:SpectralDensities}B}, presents the power spectrum of \textbf{Figure \ref{F:SpectralDensities}A} consisting of a single peak with frequency $ f = \tfrac{1}{T} $. Please note that calculating $ \ngr{\psi_f} $ the actual signal is sampled at intervals of $ \Delta x $ (equivalent to multiplying by something called \textit{Dirac comb} and the FFT also contains the frequency content of the Dirac comb produces the burst of high frequency vibrations of vibrations the right half of each stroke \cite{Blackman1959}.

\textbf{Figure \ref{F:SpectralDensities}C} is a spectrum $ \ngr{\psi_f} $ of Gaussian white noise shown in \textbf{Figure \ref{F:Gauss}A}; the ordering of panels from \textbf{\ref{F:Gauss}B} to \textbf{\ref{F:Gauss}E}, in arbitrary units. In \textbf{Figure \ref{F:SpectralDensities}C} it is observed that all frequencies have the same \comillas{weight}, that is, all frequencies contribute equally. \textbf{Figure \ref{F:SpectralDensities}D} is a spectrum \textit{Rw} of a Brownian noise $ \ngr{\psi_f} $; On the logarithmic double coordinate scale, the spectral density line is a decaying line with a slope of $ \tfrac{1}{f^2} $. \textbf{Figure \ref{F:SpectralDensities}E} is $ \ngr{\psi_f} $ of patients hospitalized each day ($ \ngr{ InP}$ in the figure) is shown in \textbf{Figure \ref{F:DailyHospitalised}A}.  

Even though the simulated sequences in \textbf{Figure \ref{F:SpectralDensities}} are only $10000 $ points and those in are only $1329 $ points long, there is an obvious similarity between the Brownian noise $ \ngr{\psi_f} $ (\textbf{Figure \ref{F:SpectralDensities}D}), and $ \ngr{\psi_f} $ of the sequence of patients in the Hospital ($ \ngr{Inp} $) between May 1, 2013 and December 19, 2017 (\textbf{Figure \ref{F:SpectralDensities}E}). In cases $ \ngr{\psi_f} $ are characteristic of the Brownian noise \textit{Rw} and none of these $ \ngr{\psi_f} $s have peaks that could indicate a suggestion of any periodic component which could some periodicity, despite short lapses or long Brownian lapses the \textit{Rw}s that could give this impression.

\textbf{Figure \ref{F:SpectralDensities}C} is a Gaussian white noise $ \psi_{f_j} $ calculated for the signal in \textbf{Figure \ref{F:Gauss}A}. In \textbf{Figure \ref{F:SpectralDensities}C} all frequencies have the same \comillas{weight}, the contribution of all frequencies is the same, \comillas{white} for this kind of noise. \textbf{Figure \ref{F:SpectralDensities}D} shows $ \psi_{f_j} $ of the Brownian noise \textit{Rw} in \textbf{Figure \ref{F:Gauss}B}, it can be seen that with logarithmic double coordinates the spectral density follows a straight line that decays with a slope of $ \tfrac{1}{f^2} $. Finally, also in \textbf{Figure \ref{F:SpectralDensities}E}, $ \psi_{f_j} $ is shown which corresponds to the sequence of daily hospitalized patients Figure \ref{F:DailyHospitalised}A. Except for the simulated sequences in \textbf{Figure \ref{F:SpectralDensities}} where $ 10^4 $ points are simulated and for Hospital data where they are 1329 points in length, there is an obvious similarity between the Brownian noise $ \psi_{f_j} $, and the $ \psi_{f_j} $ of patients discharged from the Hospital between May 1, 2013 and December 19, 2017. In both cases $ \psi_{f_j} $ are characteristic of a \textit{Rw} and \textit{none of those $ \psi_{f_j} $ has some suggestive peak of periodic components that could explain any periodicity, despite short periods in \textit{Rw}s that seem to suggest it}.

Since Mandelbrot introduced the concept of \scomillas{fractal} and published his \comillas{Fractal Geometry of Nature} \cite{Mandelbrot1967, Mandelbrot1983, Mandelbrot1989a} the notion of fractal dimension and fractal geometry have invaded virtually every corner of the economy \cite{Mandelbrot1983, Mandelbrot1997, Yuan2008, Yuan2009}, medicine \cite{Pradhan1993, Chen2005, Raghavendra2009}. Multiple corners of biology \cite{Hastings1993}, epidemiology \cite{Meltzer1991, Sevcik1998a, Sevcik2010}, physics, service systems \cite{Thomas2004, RodriguezHernandez2022}, seismology \cite{Efron1982, Cao2004, Telesca2005, Telesca2006}, the study of reversible reactions and enzymes \cite{Savageau1995, Xu2007} and much more, including so called multifrantal systems with more than one fractal component \cite{KoscielnyBunde2004, Yuan2009, Kantelhardt2008, Silva2009, Gunay2014, Krzyszczak2018, Krupienski2020}. 

\section{Parameters once relented to the fractal dimension: The Hurst exponent.}

\subsection{The Hurst water source model.}

Mandelbrot \cite{Mandelbrot1983, Hastings1993} has pointed out, that the fractal dimension is related to previous concepts associated with, for example, the flow of rivers, which would be an interesting natural phenomenon \textit{per se}. This would be the case, of the flow of the Nile River. Harold Edwin Hurst (1880 -- 1978) a British hydrologist \cite{Sutcliffe1979, Sutcliffe1999, OConnell2016, Sutcliffe2016} who was dedicated to measuring the level of the Nile river since 1906, for 62 years. Hurst published his first paper in 1951, one of the most influential and highly cited works in scientific hydrology, \cite{Hurst1951}, he was then 71 years old. Hurst\textquoteright{s} nomenclature is not easy to follow, since almost all of his variables are single letters hinting nothing about their meaning. The work of 1951 is followed by other papers in 1956 \cite{Hurst1956, Hurst1956b, Hurst1956a}.

\begin{figure}[h!]
	\centering	
	\includegraphics[width=8cm]{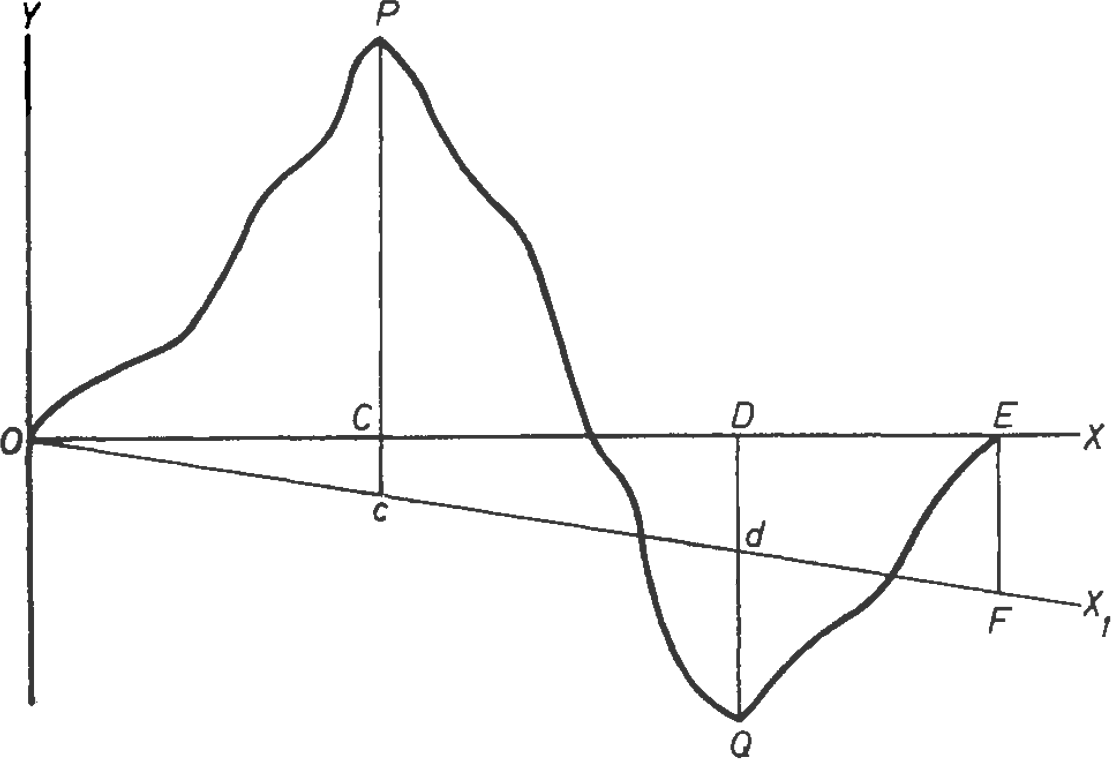}
	\caption{
		\footnotesize{
			Credits of the figure in Recognitions \cite{Hurst1956a}.
		}
	}\label{F:Hurst_Fig1}
\end{figure}

Hurst \cite{Hurst1956a} recorded the level of the Nile, since 1906 ($ \approxeq 44 $ years) preceded by about $ \geqslant 2000 $ \cite[Table 2 and Figure 1]{Hurst1956a} years of historical records. If we assume that the flow of the Nile enters a reservoir of indefinite capacity, from which there is a constant efflux equal to the annual average of discharge (evaporation, irrigation, consumption, etc.). The storage that would be required to make the discharge flow of each year is obtained by computing the \textit{continuous sums of the annual deviations from the mean}. Then the range $ R $ between the maximum and minimum of these continuous sums is either: (a) the maximum storage when there is no deficit, (b) the maximum accumulated deficit when there has never been maximum storage or (c) its sum when there is storage and deficit.

Suppose we have a record of the annual discharges of a river over a number of years, and we assume that the river flows into a reservoir of indefinite capacity, from which there exists a constant influx equal to the mean annual discharge. 

The storage required for the annual discharge to flow is obtained by adding the excess water to the average. Then the range $ \ngr{R} $ from maximum to minimum of the continuous sums is either: (\textit{\textbf{a}}), the average accumulated storage when there is never a deficit; (\textit{\textbf{b}}),  the maximum accumulated deficit when there is never a water accumulation; or (\textit{\textbf{c}}), the sum when there is accumulated and deficit water. In Figure \ref{F:Hurst_Fig1} (or  \cite[pg. 14]{Hurst1956a}), 
\begin{equation}\label{Hurst_R}
	\ngr{R= (P \text{---} C) + (D \text{---} Q)}.
\end{equation}

This is shown in \textbf{Figure \ref{F:Hurst_Fig1}} \cite[\textbf{Figure 1}]{Hurst1956a} where $ \ngr{X \in (0, X_{max})} $ is the time axis and $ \ngr{Y \in ( -Y_{min}, Y_{max})}$ is the \textit{axis of accumulated departures from the mean  $\ngr{M= \overline{Y}}$}, of $\ngr{Y}$. The curve expressing this transitions, in sequence, through the points $\ngr{0 \rightarrow P \rightarrow Q \rightarrow E }$, and $ \ngr{Q \rightarrow E} $ \footnote{$\rightarrow$ is used as \comillas{approaches} or \comillas{tends to}.}. $ \ngr{N} $  \textbf{\textit{is the number of year of observation}} \cite[pg. 14]{Hurst1956a}. During the period $ \ngr{0\rightarrow C} $ the deviations from the mean are positive and the damn storage or river level of the interval, increases from $ \ngr{0 \rightarrow P}$. Then $ \ngr{C \rightarrow D} $ period, represents a reduction in water storage in period $ \ngr{P \longrightarrow Q} $, which determines that the curve reaches a negative level with respect to the mean storage and then returns to the mean value $ \ngr{Q \rightarrow 0} $ in the $\ngr{D \rightarrow E}$, period. $ \ngr{P \rightarrow C + D \rightarrow Q = R }$, and this amount of storage would have allowed the average discharge of the period to have been maintained during it. The point $F$ whose ordinate is $\ngr{E \text{---} F = \left[ - N(M - B)\right] }$. Hurst notation is, generally speaking awkward, $\ngr{B}$ is water draft which is $\ngr{B < M}$ \cite[pg. 19]{Hurst1956a}; $\ngr{F}$ is thus an indicator of water overdraw.

\subsection{The development of the Hurst coefficient.}
Hurst \cite{Hurst1956a} usually \textit{assumed} that the flow distribution of rivers approximates a \comillas{normal} or Gaussian curve wit mean $\mu$ and variance $\sigma^2$, or $\Gaus{\mu, \sigma^ 2}$, \textit{and }that this is usually true, but that it is only part of the description of the phenomenon, since there is also a tendency to exist for high and low years that are grouped together. As far as I know, he never performed any valid statistical analysis to support this Gaussianity assumption. \comillas{\textit{Theoretical research shows that, if individual years were completely independent of each other and each year's discharges were completely independent of each other and discharges, the most likely value of $ \ngr{R \given \sigma} $ \footnote{$\given$ means \comillas{given}.} would be given by} } \cite{Hurst1956a}
\begin{equation}\label{E:Hurst_R_sigma}
	R\tfrac{R}{\sigma}  = \sqrt{\tfrac{\pi}{2}  N} = (1.25 \ldots ) \cdot \sqrt{N} ,
\end{equation}
where, as said, $ \ngr{N} $ \textbf{\textit{is the number of years of observation}} and $ \ngr{\sigma} $ is the standard deviation of dischargesfor the period considered. Experiments with random events such as flipping coins agree with this equation.

Records of discharges from a number of rivers were examined and $R$ was calculated for as many of them as were available. Unfortunately, no records were then found of discharges from a river covering more than 70 years, and therefore $\ngr{R}$ was computed for a number of rainfall records of which several spanned more than 150 years. To these were added some records of various river levels, temperatures, and pressures. A common feature of all records was that their frequency distributions, ignoring the order in which they occurred, were of the rounded type to approximate the curve of the Gaussian normal distribution. When $ \ngr{R} $ was plotted as a function of $ \ngr{N} $ the equation produced an elongated group of points
\begin{equation*}
	\tfrac{R}{ \sigma} = 1.61 \sqrt{N}
\end{equation*}
In this group there was nothing to distinguish one type of phenomenon from another, but it was clear that that $\ngr{R}$ was increasing faster than was the case with random events. This was attributed to the tendency of natural phenomena to have runs when the values as a whole are higher values and others were low. Due to the dispersion of points whose length of available records was not large enough to decide whether $ \ngr{\tfrac{R}{\sigma}} $ for natural phenomena was better represented by the square root or by some other function of $ \ngr{N} $.

In the attempt to resolve this point longer records of rainfall, temperatures, pressure, and lake levels continued to be computed for longer natural periods of natural phenomena. as a result the analysis was extended to the records of the Nilometer of Roda (Cairo, Egypt), which went back, with discontinuities, to the year 640 BC, the thickness that produced records up to 900 years, and clay tablets from which 4,000 years of record were available. 

In total, 75 different phenomena were used. In the case of three rings results from 4 different locations were considered separately, and the findings were separate, and the findings were used for each locality are the means of a group of approximately 10 trees . A record was divided into periods and for each of them the $ \ngr{\tfrac{R}{\sigma}} $ was compared. For example, with 120 years recorded it can be calculated for 3 periods of 40 years, two covering periods of 80, and a full period of 120 years. In general $ R $ was computed for periods of less than 30 years, in total 690 values of $ \ngr{\tfrac{R}{\sigma}} $ were used. A preliminary long-term examination of the data showed that $ \tfrac{R}{\sigma} $ increased faster than $ \sqrt{N} $ and less rapidly than and $ \ngr{N} $. To find the form of the relationship the statistics were divided into sets containing similar phenomena, and the sets were again divided into groups, each group consisting of a small number of values of $ \ngr{\tfrac{R}{\sigma}} $ with approximately the same value of $ \ngr{N} $. Hurst \cite{Hurst1956a} tabulates these values in his Table 1 as means of $ \ngr{\tfrac{R}{\sigma}} $ together with $ \ngr{N} $ and its logarithms. Figure 2 of Hurst \cite{Hurst1956a} presents the $ \ngr{\tfrac{R}{\sigma}} $ plotted against $ \ngr{\log(N)} $ to produce 7 good linear regressions. 

A striking point of the discussion that Hurst in all his works, is that he uses concepts such as \textit{media}, \textit{deviation from the mean}, \textit{standard deviation ($ \ngr{\sigma} $)} or \comillas{Gaussian} without any explanation or justification. It seems to mean data distributed with a pdf such as $ \ngr{\G{\mu, \sigma^2}} $, which should be distributed symmetrically around d $\ngr{\mu}$, however that symmetry does not exist in our \textbf{Figure \ref{F:Hurst_Fig1}} \cite[Original Figure 1]{Hurst1956a} or in any equivalent figure in any Hurst work. As stated above, if the median and 95\% CI are calculated for the 11 values of $ \ngr{K} $ in the in the \cite[Table 1]{Hurst1956a}, there is an asymmetry between the medians and their 95\% CI for the values of $ K $, even though we do not place the values here because the samples become small, the asymmetry also exists for the sets of $ \ngr{K} $ separating the data of $ \ngr{N = 114} $ and $ N = 25 $ from \cite[Table 1]{Hurst1956a}. This probably ignorance of Hurst and the tendency until the first half of the twentieth century to consider all Gaussian data and call them \comillas{normal}, in the sense of being the \comillas{norm}.

Hurst\s \cite[\textbf{Figure 2}]{Hurst1956a} shows that there is a linear relationship between $ \ngr{log(\tfrac{R}{\sigma})} $ and $ \ngr{\log \left( N\right)} $ for all sets in which there are enough groups. The equations of these lines are (for $\ngr{1.8 \lessapprox \log(N) \lessapprox 3} $ \cite[in \textbf{Figure 2}]{Hurst1956a})
\begin{equation}\label{E:Rel_Hurst}
	\log\left( \tfrac{R}{\sigma} \right)  = K \log\left( \tfrac{ N}{2}\right)=\log\left(R_{\sigma }\right).
\end{equation}
$ R_{\sigma}$ is called the \textit{rescaled range of fluctuations}, in this case, of the Nile \cite{OConnell2016}. The value of $ \ngr{K} $ in Table 2 of Hurst \cite{Hurst1956a} is positive and (calculated by me) $ \ngr{0.70 \; (0.69 \text{ -- } 0.72)} $ (median and 95\% confidence interval, calculated according to Hodges and Lehmann \cite{Hollander1973}), a small dispersion, perhaps somewhat skewed upwards. It is clear from the figures that for each set of phenomena a straight line fits the data well, and the remarkable fact that $\ngr{K}$, the slope of the line, varies little from one set to another. A summary of the mean of the $K$ values obtained for the 690 separate values that Hurst computed, which is equivalent to $ \ngr{\approx \tfrac{1,\bar{3}}{month}} $ of the values that Hurst measured during years in which he devoted himself to that. The fraction $ \ngr{\approx \tfrac{1,\bar{3}}{month}} $ varies little, but \textit{probably reflects what would be expected for a large river: whose changes are slow}. The number of data used $ \ngr{\approx \tfrac{1}{month}} $, is curious since some contemporary information suggests that the level of the Nile is measured daily \cite{Nilo2015}.

So far we have followed as closely as we could the data from Hurst \cite{Hurst1956a}. Here we must point out, however, that Eq. (\ref{E:Rel_Hurst}) can be rewritten in general form as
\begin{equation}\label{E:Rel_Hurst_General}
	\log_b\left(  \tfrac{R}{\sigma} \right) = \log_b\left( R_{\sigma}\right)   = K_b \log_b \left( \tfrac{ N}{2}\right) ,
\end{equation}
which suggests that straight lines are obtained by logarithms of any base. From here on we will join most authors calling $\ngr{K_b }$ the \textbf{Hurst's coefficient} and denote it as $\ngr{H}$. An extremely important particular case is
\begin{equation}\label{E:Rel_Hurst_Naeper}
	\begin{split}
		\ln\left(\dfrac{R}{\sigma} \right)   &=   \ln\left(R_{\sigma}\right) = H \ln \left( \dfrac{ N}{2}\right)  \\  
		R &= \left( \dfrac{N \sigma }{2}\right) \e^{H} \implies  R_{\sigma} = \left( \dfrac{N}{2}\right) \e^{H} \\
		\ngr{\then H } &\ngr{= \ln{\left(\frac{2R}{N\sigma }\right)} \qquad \qquad \qquad \qquad \qquad \lqqd }
	\end{split}
\end{equation}
If $R$ is Gaussian withh mean $\dot{\mu}$ and variance $\dot{\sigma}^ 2$ (that is $\Gaus{\dot{\mu},\dot{\sigma}^ 2}$, as assumed by Hurst), it may also be written
\begin{equation}\label{H_and_Gauss}
	H = \ln\left(\frac{2}{\sigma N}\right)  + \ln \left( \Gaus{\dot{\mu},\dot{\sigma}^ 2} \right),
\end{equation}
so, even when $R$ is Gaussian, $H$ wont be.

Hurst \cite[p. 19]{Hurst1956a} considers the case where the annual water demand is less than the average river flow, but not less than the minimum annual flow recorded. In this case you have an approximate, $ \ngr{B} $, equal to what is required and less than \textit{the average}, $ \ngr{M} $, and we need to know the largest \textit{accumulated deficit}, $ \ngr{S} $, which must be covered with stored water. This can be determined for any particular record of the cumulative deviation curve. 

For \textbf{Figure \ref{F:Hurst_Fig1}}, $ \ngr{0\rightarrow X} $ is the abscissa, which represents the value of the mean, with respect to which the other values of the figure are graphed. If we now produce an estimate $ \ngr{B} $ less than the average storage (axis $ \ngr{0\rightarrow X} $) it will be increased each year in the magnitude $ \ngr{M - B} $ above what was the estimated $ \ngr{M} $. This can be determined from the original curve of cumulative deviations by drawing an axis $ \ngr{0 \rightarrow X} $ passing through the point $\ngr{F} $ whose ordinate is $ \ngr{- N (M - B)} $. The curve ordinates referred to the $\ngr{0 axis \rightarrow X_1}$ show storage changes with water withdrawals $ \ngr{B} $. Storage drops from $\ngr{P}$ to $\ngr{Q}$, $\ngr{P\rightarrow c+d\rightarrow Q}$. This is the amount of storage $ S $ that would have been necessary to meet the demand $ \ngr{B} $.

$ \ngr{S} $ was determined for various water demands for each of $38$ types of phenomena such as the kinds of river discharges, rainfall, evaporation, use, and temperature that determines evaporation. The results for each class of events were grouped according to extraction, and the means of these groups were plotted. Evaporation was an important factor determining the salinization of Lake Nasser \cite{NasserLake2021}. Two relationships fit to the results really well. The adjustment is (this strange equation is Hursy\s \textit{verbatim} \cite[Eq. (3)]{Hurst1956a}), although the interrogation sign is mine:
\begin{equation}\label{E:Hurst_Eq32}
	\log_{10}\left( \frac{S}{R}\right)  = -0.08 - \ngr{\overbrace{1.00}^?} \:\left( \frac{M - B}{\sigma}\right).
\end{equation}
From there Hurst \cite[Eq. (4)]{Hurst1956a}) somehow  gets
\begin{equation}\label{E:Hurst_Eq33}
	\frac{S}{R}= 0.97 - 0.95 \sqrt{\frac{M - B}{\sigma}}.
\end{equation}
The average value of $\ngr{N}$ from which these results are calculated is $96$. It will be seen that in the adjustment, on the range of observations, there is no significant difference between one type of relationship and the other. If the figures are examined you will notice values of these kinds of phenomena are indistinguishable.

Hurst data on thr Nile river is indeed priceless, but as seen in Eqs. (\ref{E:Hurst_Eq32}) and (\ref{E:Hurst_Eq33}), his algebra is sometimes strange. Another source of uncertainty in Hurst\s data analysis is his unproven data Gaussianity, which is important to relate his exponent (now called H) with D, the fractal dimension of Nile\s water level fluctuations \cite{Mandelbrot2002, Gneiting2004, Gneiting2011, Sutcliffe2016}. For me Hurst variable denomination is also strange and not easy to follow. All these peculiarities probably reflects Hurst\textquoteright{s} isolation, and also perhaps his decision of starting to write on his work late in his life.

\subsection{Validity of the relationship between the Hurst\s coefficient  and its proposed relation with the fractal dimension.}

The origin of the association between the two parameters was initially proposed by Mandelbrot and Wallis \cite{Mandelbrot1969}, with Mandelbrot's success in finding fractal systems in the most diverse places, seemed to be evidence for a relationship between the {Hurst exponent} and the fractal dimension, $\Phi $, of two-dimensional systems was a simple relationship such as 
\begin{equation}\label{E:H_D}
	\ngr{D = 2 -H}
\end{equation}
\cite{Mandelbrot1983, Hastings1993, Sevcik1998a, Sevcik2010}. Thanks to Mandelbrot's success, fractal analysis has been applied to time series, profiles, and natural surfaces in almost every scientific discipline. 

More recent analyses \cite{Mandelbrot2002, Gneiting2004, Gneiting2011, Sutcliffe2016} indicate that the relationship between {fractal dimension and Hurst's coefficient} is actually more complex than initially thought, and is not described by Eq. (\ref{E:H_D}). 

The trouble with Hurst\textquoteright{s} analysis is that he measured Nile River \comillas{levels} respect to some river bottom oe \comillas{zero} which is bot precised. By definition the Nile River level is never negative respect to this \comillas{zero}. Because of this, the measurements in Figure \ref{F:Hurst_Fig1} are done to are some mean level which is never zero, but is taken as \comillas{zero}. And as indicated by my Eq. (\ref{E:Rel_Hurst_Naeper}) (taken from Hurst work) $H$ 

Several authors have developed independent and (more?) exact methods to evaluate the Hurst coefficient \cite{Gneiting2011, FernandezMartinez2014, Sanchez2015} \textbf{which do not presuppose the Gaussianityd of the data} that Hurst made and even consider the effect of the data \scomillas{pathological} distributed with {Cauchy PDF} \cite{Cramer1991, Pitman1993, Walck1996, Wolfram2003}. The discussion of \comillas{pathological distributions} like Cauchy's is neither trivial nor irrelevant. These distributions are not Gaussian, and have neither mean nor variance. The data that Hurst discusses are deviations from the level of the Nile from its mean, estimated $ \ngr{\bar{x}} $, this is a random variable, whose mean is
\begin{equation}\label{E;Med_difMedia}
	\overline{\Delta x_i}= \dfrac{\sum_{i=1}^{n}(x_i-\bar{x})}{n} = 0
\end{equation} 
and its apparent variance is
\begin{equation}\label{E;Var_Med_difMedia}
	\var{\overline{\Delta x_i}} = \dfrac{n+1}{n-1}\sum_{i=1}^{n}(x_i-\bar{x})^2.
\end{equation}  
If $ \ngr{x_i} $ and $R$ [see Eq. (\ref{Hurst_R})] were Gaussian as assumed by Hurst, then $ \overline{\Delta x_i} $ and $R$ would have Cauchy PDF and \textit{no operation with its apparent mean and/or apparent variance would make sense}. The Eqs. (\ref{E;Med_difMedia}) and (\ref{E;Var_Med_difMedia}) can be calculated but are meaningless for a Cauchy distribution without central moments \cite{Grassberger1981, Grassberger1983, Grassberger1983a, Grassberger1983b, Theiler1990, Grassberger2004}. If $ \ngr{x_i} $ is not Gaussian the problem is similar, an unknown pathological random variable whose mean and variance estimated from a sample, are meaningless, because the pdf of count come has no mean or defined variance. \textit{All sampling theory makes sense if and only if, the parameters that are estimated from the sample are approximations of the parameters of the population we sample}.

Recent evidence indicates that the relationship between the Hurst exponent and the fractal dimension is neither linear nor simple. \textit{It does not seem justified to consider that the fractal dimension and the Hurst exponent are related to a simple linear relation}, so we will not devote more attention here.

\section{Chaos, Strange Attractors, and Mathematics.}

The idea that certain systems, numerically apparently very simple, could not have predictable values except at short times, was born when Lorenz \cite{Lorenz1963} tried to solve a model of climate. The Lorenz model focused on three simultaneous equations (with slightly modified notation) which ate a numerical approximation to Rquation System (\ref{E:Lorenz}):
\begin{equation}\label{E:Lorenz_Sis}
	\begin{matrix}
		x_{t+1} &=& -\sigma x_t + \sigma y_t \\
		y_{t+1} &=& x_t z_t + \gamma x_t - y_t\\
		z_{t+1} &=& x_t y_t - b z_t
	\end{matrix}
\end{equation}
where $ x_t$ , $y_t$ and  $z_t $ represent the model variables in the iteration $ t $, and $ \sigma $. and $ \gamma$ are parameters set by the researcher before simulation (see \cite{Lorenz1963} for details). In 1963 Lorenz used an analog computer where the variables of the simulated model were introduced by varying knobs that modified parameters of the simulator circuit. The analog circuit produced electrical signals that represented the results, these were printed, and before turning off the computer, Lorenz saved the values of the last set $ \{x_{t+1}, y_{t+1}, z_{t+1} \}$ that he reintroduced into the system when he turned on the device the next time.

As long as the system operated without interruptions, the values stayed within a parallelepiped rectangle of sides $ \ngr{\{X_{max}, Y_{max}, Z_{max}\}} $ (a set of its solutions is presented in \textbf{Figure \ref{F:Lorenz}}). The solution never passes through the same point twice, but evolves around that right parallelepiped forever. It seems as if something attracts the system to evolve there, thus the concept of \textbf{\textit{attractor}} was coined to describe situations like this. Simply put, a {attractor set} \textbf{A} for a dynamical system is a closed subset of {phase space} \cite{StateSpace2008} such that \comillas{many} (most?) of the initial conditions the system evolves towards \textbf{A} \cite{Attractor2006}. In the case of the Lorenz ensemble, the system solution never passes through the same place twice, a feature that coined the name \textit{\textbf{strange attractor}}. In the Lorenz model the impossibility of reproducing or continuing from the point where it ended. This is due to the impossibility of introducing an earlier trajectory due to the imprecision of the controls of analogue systems. With digital computers, the researcher chooses an initial numerical value of his interest, this initial value is \comillas{type} always the same. At least on the same computer with the operating system and the accuracy of the processor it has, it is always handled the same and the initial (graphic) calculation is always the same. But, if the system stops and its variables are saved, behave the same as the analog  Lorenz's computer, the graph when rebooted does not follow the same trajectory: it becomes irreproducible: chaotic.

For nearly 20 years, Lorenz's work received little attention. Only when strange attractors began to be observed in physics and then in almost every branch of science did Lorenz's work became fundamental. Perhaps because of the shape of the Lorenz Attractor, the sensitivity to initial conditions in climate prediction was called the \comillas{butterfly effect}. Strange attractors appeared everywhere \cite{StrangeAttractor2014}. The mathematics behind strange attractors can be extremely simple as follows from the set of equations (\ref{E:Lorenz_Sis}), but the implications deep.

\begin{figure}[h!]
	\centering	
	\includegraphics[width=8cm]{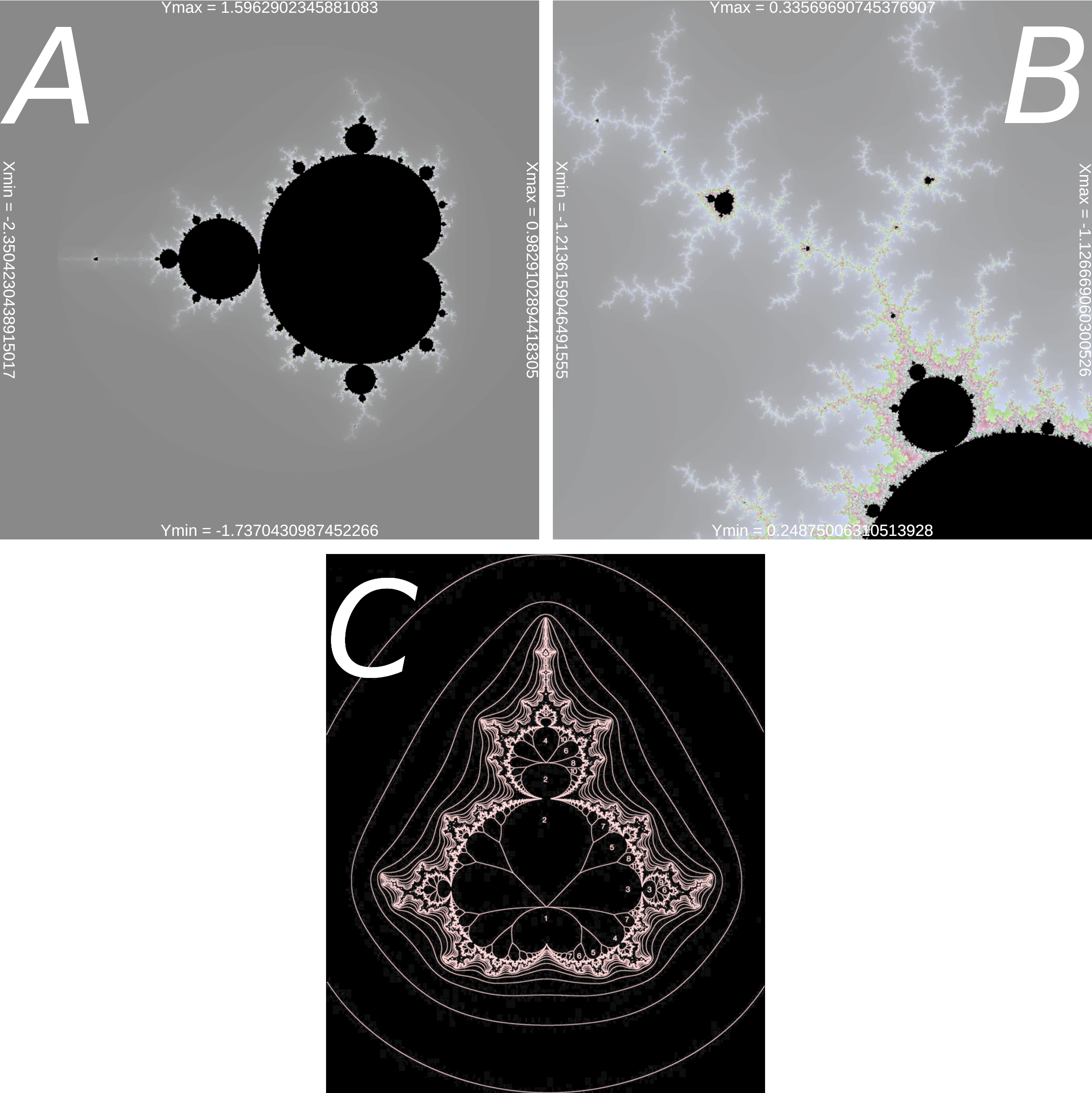}
	\caption{
		\footnotesize{
			\textbf{Mandelbrot set: Calculation in relation to $ \ngr{n} $.} \textbf{Panel A:} With initial values to display \comillas{all} the set. Panel A was calculated by taking up to $ n = $1024. Images calculated with JavaLab\s Java program, available at \cite{MandelbrotSet2020}. The scales indicated in \textbf{Panel A} are: $ x_{max} = 0.98291028944183 $, $ x_{min} = 0.98291028944183 $, $ y_{max} = 1.59629023458811 $ and $ y_{min} = -1.73704309874523 $. \textbf{Panel B} are: $ x_{max} = -1.12666906030052 $, $ x_{min} = -1.21361590464915 $, $ y_{max} = 0.335656969074538 $ and $ y_{min} = 0.248750063105139 $. The values of $ \left( x_{min}, x_{max}\right) $ and $ \left( y_{min}, y_{max}\right) $, are the initial values of the calculation in each case: $ Z_o = x_{min} + x_{min} \J $ and $ C = y_{min} + y_{max} \J $- \textbf{Panel B:} Enlarged to show the appearance of structures similar to those in Panel A, as it is linearly expanded $ \approxeq 38.4 $ times. \textbf{Panel C:} Based on Figure 34. by Mandelbrot \cite{Mandelbrot1986}. In Panel C you can see some digits that are the $ n = 1, \ldots, 7$, of a mapping such as $ \ngr{Z_{n+1} \mapsto Z_n^2 + C} $. The figure, which shows self-similarities when enlarging as black images in the branches of the whole. The scales are the coordinates of the graph.
		}
	}\label{F:Mandelbrot_Set1}
\end{figure}

\subsection{Fractals and two-dimensional strange attractors.}

As we have already said, the definition of fractal by Mandelbrot \cite{Mandelbrot1983, Mandelbrot1989a} occurred almost 20 years after Lorenz\textquoteright{s} work \cite{Lorenz1963} and introduced interest in a class of, especially, two-dimensional strange attractors (see \textbf{Figure \ref{F:Mandelbrot_Set1}}). They were not really new, the sets of Julia \cite{Julia1922} were published 60 years earlier. Just as the development of nonparametric statistics benefited from the emergence of computers after the 1940s, and their cheapening increased accessibility in the 1950s and cheapening with powerful and easily accessible microcomputers in the 1980s. As consequence, rapid advances occurred in processioning of iterative calculation and graphing. Thus systems that require iterative computation and graphing were developed. Such as systems, representing chaos, iterative and fractal systems such as the Julia sets \cite{Julia1922, Mandelbrot1966, Lei1990} and the beautiful and fascinating Mandelbrot set. \cite{Mandelbrot1980, Mandelbrot1986, Mandelbrot1986a, Mandelbrot1989, Mandelbrot1989a, Lei1990}.

The initial problem with fractal analysis was that physicists used it to study diverse systems where they determined the fractal dimension with the same precision with which other physical constants are studied. 

\subsection{The Mandelbrot set.}

The Mandelbrot set \cite{Mandelbrot1966, Mandelbrot1986, Mandelbrot1986a, Branner1989}the graphs shown in \textbf{Figure \ref{F:Mandelbrot_Set1}}] is another example of an attractor as simple as Eq. (\ref{E:Lorenz_Sis}) by Lorenz \cite{Lorenz1963} which, however, hides a great conceptual depth. 

Let's look at some details about the Mandelbrot equation, it is an equation that has solutions based on a variable $ Z \in \Com$ and a constant $ C \in \Com $, as follows:
\begin{equation}\label{E:Mandelbrot_set}
	\begin{split}
		Z_{n+1} &= \left\langle Z_n^2 +  C \right\rangle \ngr{\given \left[ \left(  Z_n= z_{n,re} + z_{n,im} \J\right) \land \left( C= c_{re}  + c_{im} \J \right) \right]} \\
		& \implies  \left( Z_{n} \in C \right)  \land \left( Z_{n+1} \in C\right)  \land \left( C \in C \right) 
	\end{split}
\end{equation}
where $ \J = \sqrt{-1} $. Eq. (\ref{E:Mandelbrot_set}) is sometimes written \cite{Mandelbrot1980} $ Z_{n+1} \mapsto Z_n^2 + C $. With the $Z mapping \mapsto f(Z, C)$. Where $ f \in \R $ is an \textit{irrational function ($\Q$)} so that $ (f(Z) \in \Q) $ of $ Z $ and $ C $, consider the iterated maps $ Z_n = f(f( \ldots f(Z_0) \ldots)) $ of the point $ Z_0 $.

Square the complex numbers $Z$and $C$ to create a new number $Z$. Square the new number $ Z $ and add $ C $ to produce another $ Z $. Mandelbrot's set is another example of mathematical simplicity that hides a great depth of concepts. The Eq. (\ref{E:Mandelbrot_set}). apart from including complex numbers, $ Z \in \Com $, can hardly be simpler, it only requires a square of a complex number
\begin{equation}\label{E:Zn}
	\begin{split}
		Z_n² &= \left[ \left( z_{re,n}^2 - z_{im,n}^2\right)  + 2 \left( z_{re,n} z_{im,n}\right)\J \right]
	\end{split}
\end{equation}
Notice that there is a particular case where $ Z_n^2 $ becomes imaginary, this is
\begin{equation}\label{E:Zparticular}
	\begin{split}
		z_{re,n} &= z_{in,n} \implies Z_n^2= 2 \left( z_{re,n} z_{im,n}\right)\J \\
		\ngr{\then Z_n} &\ngr{= \J \sqrt{2 \ z_{re,n} z_{im,n}}= \sqrt{2}} z_{.n}\; \J
	\end{split}
\end{equation}
The dot in $ z_{.n} $ recognizes that both components are equal. But in general, if the situation foreseen in Eq. (\ref{E:Zparticular}) holds, the next real component of $ Z_{n+1} $ depends entirely on the complex constant $ C $:
\begin{equation}\label{E:Zn+1}
	\begin{split} 
		Z_n^2 &= z_{re,n}^2 - z_{im,n}^2  + 2 \left( z_{re,n} z_{im,n}\right)\J  \\
		Z_{n+1}^2 &= z_{re,n}^2 - z_{im,n}^2  + 2 \left( z_{re,n} z_{im,n}\right)\J  + c_{re}  + c_{im} \J\\
		\cdots &=  \left[ z_{re,n}^2 - z_{im,n}^2 + c_{re}\right]  + \left[  2 \left( z_{re,n} z_{im,n}\right) + c_{im} \right] \J \\
		\ngr{\then Z_{n+1}} &\ngr{= \sqrt{ \left[ z_{re,n}^2 - z_{im,n}^2 + c_{re}\right]  + \left[  2  z_{re,n} z_{im,n} + c_{im} \right] \J } \quad \lqqd}.
	\end{split}. 
\end{equation}

Repeat this $ \aleph_0 $ times \cite{Branner1989, MandelbrotSet2020, MandelbrotSet2022}\footnote{$\aleph_0$ is an infinitely large cardinal number}. A graph, co,or those presented here in \textbf{Panel \ref{F:Mandelbrot_Set1}A} is the solution \comillas{complete} and \scomillas{continuous} of the Eq. (\ref{E:Mandelbrot_set}) and the \textbf{Panel \ref{F:Mandelbrot_Set1}B} is a \comillas{window} of a solution which extends $ \approxeq \times 40 $ iterations \cite{Branner1989, MandelbrotSet2020, MandelbrotSet2022}. 

\subsubsection{How to calculate the Mandelbrot set.}

Mandelbrot found that the value of $\ngr{Z}$ continued to increase or oscillate between two smallvalues  \comillas{depending on $ \ngr{C }$)} \cite{MandelbrotSet2020}. He used a computer to calculate each value of $ \ngr{C} $ and plotted it on the screen as a not \comillas{radial} dot $ \ngr{Z }$. The result is somewhat difficult to describe, a purely generic structuring process (squares, triangles, circles), each Mandelbrot graph is the solution of Eq. (\ref{E:Mandelbrot_set}) up to $ \ngr{n+1 \rightarrow \aleph_0} $. The resulting graph is called the {Mandelbrot set}. Mandelbrot continued to magnify the image, but the same structures continued to appear (self-likeness). No matter when he left, for $ \ngr{n+1 \rightarrow \aleph_n} $, the same images continued to appear.

\begin{figure}
	\centering
	\includegraphics[width=6cm]{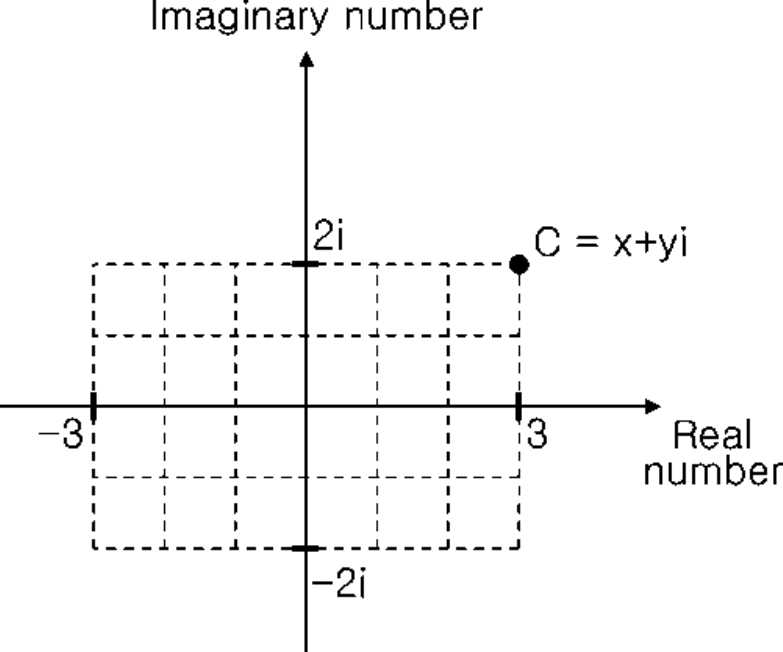}
	\caption{
		\footnotesize{
			\textbf{Mandelbrot set. Determination of $ \ngr{C} $}. A window defined for complex numbers $ \ngr{(x,y}\J)$ with $ \ngr{C} $ defined in the upper right corner. Figure based on another of  \cite{MandelbrotSet2020}.
		}
	}\label{F:MandelbrotCS4}
\end{figure}

The procedure can be described as: create a window with complex coordinates ($ x, y \J $) with one pixel per computer screen number and call it $ C = x, y \J) $, the others are possible values of $ Z_{n+1} \mapsto Z_n^2 + C $ defined in Eq. (\ref{E:Mandelbrot_set}). Usually, the range of the $ x $ axis goes between $ (-3 \leftrightarrow +3) $, and the range of $ y $ values goes between $ (-2 \leftrightarrow +2) \J $. In this range, you can see the entire picture of the Mandelbrot set. \textit{if the values of $ Z $ do not diverge} is \textit{represented on site with a black dot} on the screen, the result is a Mandelbrot set. Usually if $ n > (2 \leftrightarrow 4) $ this is greater than, it is recorded as a {divergence} and the recurring relationship is stopped.

\begin{figure}[h!]
	\centering
	\includegraphics[width=10cm]{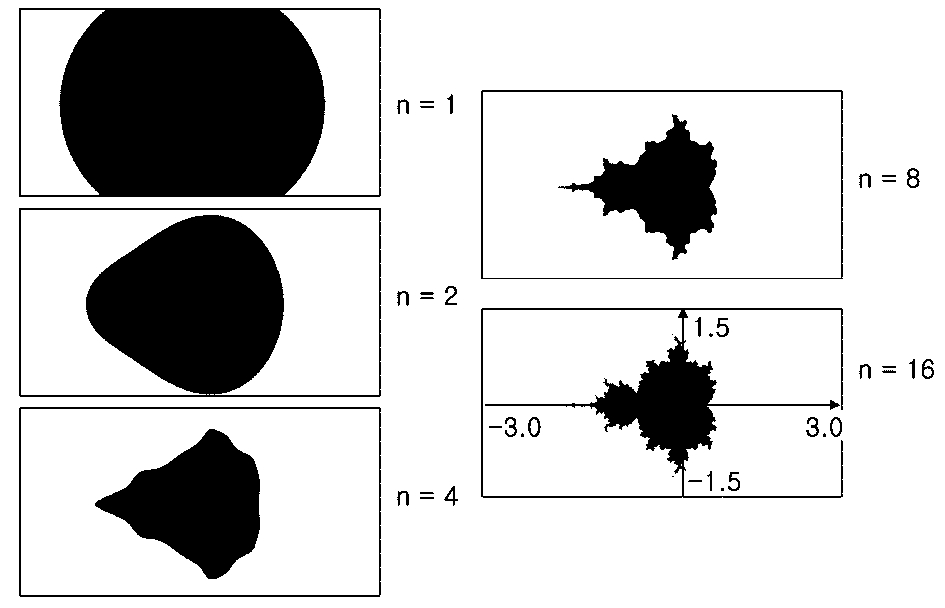}
	\caption{
		\footnotesize{
			\textbf{A window defined 	for complex numbers $ \ngr{(x,y\J)}$ with $ C $ defined in the upper right corner}. Figure based on another of $ \ngr{n}$  \textbf{the number of iterations}. Figure based on \cite{MandelbrotSet2020}
		}
	}\label{F:MandelbrotCS5}
\end{figure}

The more the recurring relationship is repeated, the more detailed we can get a figure. However, this cannot be calculated indefinitely. Due to limitations of computing power, even if high-precision real numbers such as those indicated in \textbf{Figure \ref{F:Mandelbrot_Set1}} are used, therefore the process stops when it reaches a reasonable precision: $ n=1024 $ in the \textbf{Figures \ref{F:Mandelbrot_Set1}A} and \textbf{B}, or $ n=16 $ in the \textbf{Figure \ref{F:MandelbrotCS5}}.

There is an aesthetic aspect to fractal images. An artistic look can be added to graphics by adding arbitrary colors to successive $Z_n$ values. This is done, sometimes associating color regions to ranges of values of $ n $, where usually the use of black for $ n \leqslant 7 $ is maintained, and then other color bands to the liking of the calculator. One of many examples are our \textbf{Figures \ref{F:Mandelbrot_Set1}A}, \textbf{B} and \textbf{C} The graphics using the Mandelbrot set have become an element to show aesthetics and computing power accompanied by the sets of Julia \cite{Peitgen1986}.

Some details of the calculation and graphing of the Mandelbrot set have been extended here to describe in part the information contained in the Mandelbrot set, the fruit of human ingenuity and the simple mathematics that underlies it.

\section{Use and abuse of the fractal dimension.}

The introduction of the concept of {chaos} in climate by Lorenz \cite{Lorenz1963, Lorenz1969, Palmer2022}, from where, after a delay, it invaded the rest of science widely, and the introduction of the concept of {fractal} by Mandelbrot \cite[pg. 15 and Ch. 39]{Mandelbrot1983} changed the view of the science of numerous systems, as it emerges, partially, from the previous discussion. His problem: the same as data analysis and statistics: mathematics.

\subsection{Misuse of the fractal dimension estimation methods.}

A fundamental problem in fractal analysis is using a poor method or not understanding the limitations of the method being used. In this paper we have presented Mandelbrot\s original definition of the concept of fractal dimension.

According to Mandelbrot \cite[pg. 15 and Chap. 39]{Mandelbrot1983}
\begin{quote}
	\textquotedblleft A fractal is by definition a set for which the Hausdorff{-}Besicovitch dimension strictly exceeds the topological dimension. Every set with a non integer \textit{D} is a fractal.\textquotedblright
\end{quote}
The definition has a problem, however, since it depends on estimating the Hausdorff--Besicovitch dimension of the interest function. 

Although Hausdorff--Besicovitch definition is simple:
\begin{equation*}
	D_{HB}=-\underset{\epsilon \rightarrow 0}{\lim }{\frac{\ln [N(\epsilon )]}{\ln (\epsilon )}} = -\underset{\epsilon \rightarrow 0}{\lim } \log_{\epsilon}[N(\epsilon)]
\end{equation*}
the definition implies that $ N(\epsilon) \rightarrow \ \infty $, a number that, however large, in most experimental cases is not available.

In the definition of $ N(\epsilon) \implies N(\epsilon) \in \R $ a very large real number, $ \approx \infty $, but this may not seem true in waveforms sampled in discrete mode.

A second interpretation may be that $ \left\lbrace N_i(\epsilon) \right\rbrace$ represents a set of size $ N $ such that 
\begin{equation}\label{E:NatOf_i}
	N_i(\epsilon) \mapsto 
	\begin{cases} 
		i \in \{1 \leftrightarrow \aleph_0\} \implies N_i(\epsilon) \in \R\\ 
		i \in \{1 \leftrightarrow \infty\} \implies N_i(\epsilon) \in \R 
	\end{cases} ,
\end{equation}
but this may not be true, in the case of, for example, the decimal digits of $ \pi $: $ N_i(\epsilon) \in \{0,1,\ldots,9\} $ \cite{Sevcik2017b}. In such cases, $ D_{HB} $ may be undefined, or greater, than the value of $ D $ for $ \left\lbrace \log_\epsilon \left[ N(\epsilon) \rightarrow \aleph_0 \right] \right\rbrace $. Under these condition Sevcik\s $D_S$, or any empirical estimate of $ D_{HB} $ will be partially uncertain. In Eq. (\ref{E:NatOf_i}), $\aleph_0$ is the larges value of $i$ which is $(i \in \Z)$, the set of all integer numbers. Minimizing that uncertainty is represented by a very large set of numbers. If $\aleph_0 \rightarrow \infty$ the uncertainty of the decimal of $\pi$ will be represented by real numbers \cite{Sevcik2017b}

\subsection{Accurate ways to messure $D$.}

The most accurate (for even  $n > 2$ dimensional spaces) ways to determine $ D $ were introduced by physicists \cite{Newhouse1978, Theiler1990, Grassberger1981, Grassberger1983, Grassberger1983b, Newhouse1978, Grassberger2004}. This often involves studying the same system in Euclidean spaces of different dimensions, using Lyapunov exponents \cite{LyapunovExponent2022}, and demands the use of complicated mathematics. 

\subsubsection{Baseless relation between fractal dimension and the Hurst\s coefficient ($H$).}

In the early days of the definition and \comillas{extension} of the concept of {fractal}, Mandelbrot \cite{Mandelbrot1983} was {liberal}. The association between the {Hurst exponent} and the {fractal dimension} that we discuss in this here, we find an example.

\subsection{Katz \comillas{fractal dimension} does not measures fractal dimension.}

\begin{figure}[h!]
	\centering
	\includegraphics[width=12cm]{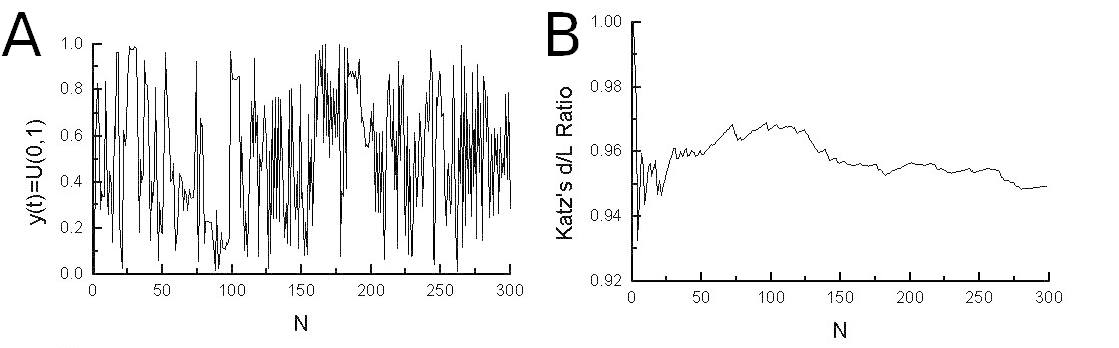}
	\caption{
		\footnotesize{
			\textbf{Properties of the ratio $ \ngr{d/L} $ for a curve generated with Monte Carlo simulation \cite{Dahlquist1974}}. \textbf{Panel A:} Shows a waveform constructed by joining with straight lines a series of uniform pseudo random numbers in the interval $ (0,1) $. \textbf{Panel B:} Ratio $ d/L $ calculated for the corresponding point of the abscissa in the top panel. In both panels the abscissa is $ N $. other details in the original publication \cite[Figure 2]{Sevcik1998a,Sevcik2010} and in the text of the this paper.
		}
	}\label{F:Simd/L}
\end{figure}

Figure \ref{F:Simd/L}B shows that $d/L$ oscillates a lot at first, but then stabilizes with $ N \geqslant 50 $. This confirms the uselessness of [Eq. (\ref{E:D_K})] to determine the fractal dimension of a curve. This is true for all waves for which ${d/L}$ is asymptotically constant after sampling ${m < \infty}$ points for ${N' < \infty}$ Eq. (\ref{E:D_K_Errada}) also implies that the values of $D_K$ is arbitrarily determined by the choice of $N$. The boundary condition, ${N'\rightarrow \infty}$, can be made if ${t_{max}}$ (the sampling duration) remains constant but the sampling interval ${\delta \rightarrow 0}$. I submitted Eq.(\ref{E:D_K_Errada}) in the first version of my papr \cite{Sevcik1998a, Sevcik2010} to the journal whee Katz paper appeared and was rejected, more than 10 years later the published a Letter to The Editor with an \comillas{experimental proof} suggesting that there is \comillas{something wrong} with $D_K$ \cite{Castiglioni2010}. In spite of all this $D_K$ is still used by authors with ignorance of algebra, usally because they know nothing else and they found it \scomillas{pretty} \cite{Dubuc1989, Embrechts1993, EblenZajjur1999, Ripoli1999, Esteller2001, Purkait2003, Gnitecki2004, Gnitecki2005, Gob2005, Pan2008, Raghavendra2009, RamirezVazquez2014, MorenoGomez2020, Hadiyoso2021, Gil2021}. 

Katz \cite{Katz1988} Eq. (\ref{E:D_K}) is wrong \cite{Sevcik1998a, Sevcik2010} This becomes apparent if we calculate the following limit:
\begin{equation}\label{E:D_K_Errada}
	\ngr{\underset{N' \rightarrow \aleph_0}{\lim} D_K} = \underset{N' \rightarrow \aleph_0}{\lim} \left[ \frac{\log (N')}{\log (N')+\log (d/L)}\right]   = \ngr{1 \qquad \lqqd}.
\end{equation}
In the original version \cite{Sevcik1998a,Sevcik2010} the limit is made up to $ \infty $, but since $ N' \in \Z $, it seems here more rigorous to do it up to $ \aleph_0 $. The limit in Eq. (\ref{E:D_K_Errada}) shows that Eq. (\ref{E:D_K}) predicts that all curves are analyzed with the Eq. (\ref{E:D_K}) as the discretization is improved by making $ N' $ very large or $ \aleph_0 $, we will discover that all curves are straight lines with $ D_K= 1 $.  
It could be argued that the limit in Eq.  (\ref{E:D_K_Errada}) has been derived assuming that $ d/L $ is a constant. This is discussed in \cite{Sevcik1998a, Sevcik2010} with the analysis of data produced with Monte Carlo simulation \cite{Dahlquist1974} as shown in Figure \ref{F:Simd/L}.

Also a \scomillas{liberal} idea of Mandelbrot \cite{Mandelbrot1983} was associating complex systems with the fractal dimension. An example of this was its association of the length of a river with a straight line distance between its source and its mouth. This view inspired the fractal dimension estimated with Katz's method \cite{Katz1988}:
\begin{equation*}
	D_K =\frac{\log (N')}{\log (N')+\log (d/L)} . 
\end{equation*}
This definition of Katz has a problem: \textit{\textbf{only applies to short waveforms}} and \underline{in} \underline{no} \underline{case} \underline{converges} \underline{to} $\ngr{\Phi }$. This is easy to see considering the limit in Eq.(\ref{E:D_K_Errada}) \cite{Sevcik1998a, Sevcik2010}.

\subsection{Sevcik\s fractal dimrnsion.}

Given the inability of $ D_K $ to predict $ \Phi $, another method was developed to calculate the fractal dimension \cite{Sevcik1998a, Sevcik2010}, this method describes what is now called Sevcik's fractal dimension in GogScholar \cite{GoogleScholar2020} refers \RefDS (\printdayoff\today) citations of the use of Sevcik\s fractal dimension \cite{Sevcik1998a, Sevcik2010}, using this algorithm. The \cite{Sevcik1998a, Sevcik2010} algorithm is a double linear transformation [Eqs. (\ref{E:Sevcik_x_tranf}) -- (\ref{E:VarDelta_y_LN})].

As said $D_S$ is calculated in a linear transformation of the space where the waveform exist, thus both spaces have the same metric properties, whih includes $D_{HB}$, $\Phi$ and $D_S$ \cite{Barnsley1993, Sevcik1998a, Sevcik2010} \textit{\textbf{if}} $\ngr{N\longrightarrow \infty}$. Since the topology of a metric space does not change under a linear transformation, it is convenient to transform a wave into another normalized space, where all axes are equal. I proposed the use of two linear transformations that map the original wave function into another embedded in an equivalent metric space [please see \cite{Barnsley1993} for a very readable discussion of metric spaces]. In recent years, this way of calculating the fractal dimension of a wave has been called the {Sevcik fractal dimension} \cite{RodriguezHernandez2022}, which here we call $ D_S $. This determines that the original function and is transformed into anther into a space with equal coordinates and with units. \textbf{Thus it is not surprising that, as shown in Subs-subsection \ref{S:Convergence}, $\ngr{D_S}$ converges to $\ngr{\Phi}$ as $\ngr{N \rightarrow \infty}$.}

Slil if
\begin{equation*}
	N \rightarrow  
	\begin{cases}
		\aleph \in \Z\\\
		\infty \in \R
	\end{cases},
\end{equation*}
$N$ is \textbf{VERY} large, and nobody works that much,so, for any real life estimate of $D_S$ we will have  $D_S < \Phi $. If we are interested in the decimal sequence of $\pi$ \cite{Sevcik2017b}, we can compare this sequence,  $\{i_{\pi,i}\}_{i=1,2,\ldots,k,\dots N} =\{i,_{\pi,1}, i_{\pi,2},\ldots, i_{\pi,k},\ldots  i_{\pi,N} \} \land \left[i_{\pi,k} \in \{0, 1,2,3,4,5,6,7,8,9\} \}\right] \in \Z $. If the decimal sequence of $\pi$ is bob periodic and all digits are equally likely, they must be distributed as a random uniform variable \cite{Guttman1965} set where the probability of any digit in the set $\{0, 1,2,3,4,5,6,7,8,9\}$ has a probability of $\frac{1}{10}$,i.e., with a \comillas{reasonable} $N$: a sort of white noise \cite{Hastings1993}. A real white nose is shown in Figure \ref{F:Gauss}A, where the points are all Gaussian real numbers.

As of July 2024, $Pi$ has been calculated to $2.02 \cdot 10^{14}$ digits \cite{Pi2024a}, and $10^9$ are easy to generate on a personal computer \cite{Bellard2010, Sevcik2017b}. Thus many sequences of series of random uniformly distributed each with a probability $\frac{1}{10}$ decimal digits $N$ and compared with the same numbers of $\pi$ of decimals \cite{Sevcik2017b}. The same procedure may be used to compare an observed series and compare ot with a random sequence of known properties.

\subsection{Avoiding intuitive conclusions in relation with $\Phi$.}

Although it is possible to prove for fractal functions such as {Koch snowflake} \cite{Koch1904, Sevcik1998a, Sevcik2010} that $ \underset{N \rightarrow \aleph_0}{\lim} D_S = D_{HB} $, for more complex functions this can be difficult \cite{Sevcik1998a, Sevcik2010, Sevcik2017b, RodriguezHernandez2022} ka demonstration may not be easy and require \textbf{\textit{relianle}} Monte Carlo simulations \cite{Shi2012, Montero2020, Sevcik1998a, Sevcik2010, Sevcik2017b, RodriguezHernandez2022}.

Intuitive estimations of $\Phi$ are dangerous. Even a solid mathematician as Mandelbrot devised more or less intuitive estimates for rivers using the length of the river and the straight line distance between the river fountain to the river end \cite{Mandelbrot1983}, the Katz $D_K$ fractal dimension is a generalization Mandelbrot\s river $\Psi$. Even if the fractal dimension defined by Mandelbrot is the ratio of two \comillas{natural} river, perhaps better for long rives, but $D_K$ gets worse as the number of discretized waveform points grows \cite{Katz1988, Sevcik1998a, Sevcik2010}.

\section{Pitfalls using the fractal concept.}

\subsection{The use of the term \scomillas{fractal} to promote unscientific theories.}

The term fractal is sometimes used as as manner to provide a \scomillas{deep scientific} meaning, to propositions lacking any scientific sport. This seems the case this seems the case of a theory unifying gravity and quantum physics \cite[among several YouTube videos and a LinkedIn comunication]{McGinty2023, McGinty2023c} or with fancy sounding theories to synthesize drugs \cite{QuantumFractals2024}. 

\subsection{Cautions when using fractal dimension estimates.}

Authors make mistakes and reviewers not always understand the concept of fractal dimension. Between the non physicists, fractal analysis became fashionable in this century. Unfortunately, too many authors (and their reviewers in journals) do not pay attention to the mathematics of the fractal analysis methods, thus the Katz and Higuchi methods have gained undue popularity, specially in medical, psychological, psychiatric an low level technique papers, there are too many examples to cithem precisely here but many examples can be found with GoogleScholar (\url{https://scholar.google.com}) or with sites such as Academia.edu (\url{https://www.academia.edu}) (among other scientific paper browsers) intermixed with valuable papers.

Another problem is (excuse the neologism) that we will call: the \comillas{curve nicety}: a principle that demands that, above any other principle, the curves and data that we enter in a work must look \scomillas{pretty}. Of course, if they are also true better than better. While Eq. (\ref{E:D_K_Errada}) \cite{Sevcik1998a, Castiglioni2010, Sevcik2010} demonstrates beyond doubt that the called \comillas{fractal dimension of Katz} \cite{Katz1988} is different from $1$ only if the starting points of any waveform are considered and is equal to 1 for any waveform other than a straight line when $ N \rightarrow \aleph_0 $ or $ N \rightarrow \infty $ and $ \underset{N\rightarrow \infty }{\lim}D_{HB} \neq 1.0 $. The problem is that Eq. (\ref{E:D_K}) does not have an \comillas{optimal} to estimate $ D_K $ Eq. (\ref{E:D_K_Errada}), \textbf{\textit{always underestimates $\Phi$}} even if you underestimate it more with large $ N $. However a number of authors base their choice of a method of assigning fractal dimension to their curves choose the one that makes their data look \comillas{prettier} and thus preserve meaningless methods \cite{Dubuc1989, Embrechts1993, EblenZajjur1999, Ripoli1999, Esteller2001, Purkait2003, Gnitecki2004, Gnitecki2005, Gob2005, Pan2008, Raghavendra2009, RamirezVazquez2014, MorenoGomez2020, Hadiyoso2021, Gil2021,Anier2004, Doyle2004, Klonowski2005, Spacic2011, Spasic2011a, Harne2014, Alnuaimi2017, Cukic2018, Liehr2019, Liehr2019a,  Shamsi2021, HiguchiDimension2022}. 

\section*{Acknowledgments.}

\subsection*{Free open source software used.}
This manuscript was written in \LaTeX{ }({\TeX}Live 2023, \url{https://www.tug.org/texlive/}).Tex using \textit{{\TeX}studio} \TEXs{ }(\url{https://www.texstudio.org})for Linux, an open source free \LaTeX{ }editor. All the silverware used here ran under Cinnamosn Linux Mint 22 \comillas{Wilma} (\url{ttps://www.linuxmint.com/download.php}).

\subsection*{Figures reproduced fron the literature.}
\textbf{Figure \ref{F:Hurst_Fig1}} of this work is the reproduction of the \textbf{Figure 1} of the work \cite{Hurst1956b} published on line by \textit{Ministry of Public Works}, Egypt on January 04, 2010 \cite{Hurst1956b} of work \cite{Hurst1951}. 

The article is provided under license from Taylor \& Francis which the work states as follows:  
\begin{quote}
This article may be used for research, teaching, and private study purposes. Any substantial or systematic reproduction, redistribution, reselling, loan, sub-licensing, systematic supply, or distribution in any form to anyone is expressly forbidden. Terms \& Conditions of access and use can be found at http://www.tandfonline.com/page/ terms-and-conditions. \footnote{Internet access reproduced here exactly from the original.}
\end{quote}

\medskip


\end{document}